\documentclass[aps,pre,onecolumn]{revtex4}
\usepackage{graphicx, bm, bbm,amsmath, amssymb, epsfig}

\pdfoutput=1

\newcommand{\nn}{\noindent}

\newcommand{\bq}{\begin{align}}
\newcommand{\eq}{\end{align}}

\begin{document}
\title{Efficient sliding locomotion of three-link bodies}
%\title{Efficient sliding locomotion of three-link bodies in resistive environments}

\author{Silas Alben$^*$}
%\thanks{}
\affiliation{Department of Mathematics, University of Michigan,
Ann Arbor, MI 48109, USA}
\email{alben@umich.edu}

\date{\today}

\begin{abstract}
We study the efficiency of sliding locomotion for three-link bodies in the presence of dry (Coulomb) friction.  Friction coefficient space can be partitioned into several regions, each with distinct types of efficient kinematics. These include kinematics resembling lateral undulation with very anisotropic friction, small-amplitude reciprocal kinematics, very large amplitude kinematics near isotropic friction, and kinematics that are very asymmetric about the flat state. In the two-parameter shape space, zero net rotation for elliptical trajectories occurs mainly with bilateral or antipodal symmetry. These symmetric subspaces have about the same peak efficiency as the full space but with much smaller dimension. Adding modes with two or three times the basic frequency greatly increases the numbers of local optimal for efficiency, but only modestly increases the peak efficiency. Random ensembles with higher frequencies have efficiency distributions that peak near a certain nonzero value and decay rapidly up to the maximum efficiency. A stochastic optimization algorithm is developed to compute optima with higher frequencies. These are simple closed curves, sharpened versions of the elliptical optima in most cases, and achieve much higher efficiencies mainly for small normal friction. With a linear resistance law, the optimal trajectories are similar in much of friction coefficient space, and relative efficiencies are much lower except with very large normal friction.
\end{abstract}


\maketitle

\section{Introduction}

In this work we investigate sliding locomotion by three-link bodies. Such bodies are a benchmark system for studying the basic physics of locomotion, for swimming microorganisms \cite{Pu1977a,BeKoSt2003a,TaHo2007a,AvRa2008a,RaAv2008a,HaBuHoCh2011a,huber2011micro,passov2012dynamics,alouges2013self,giraldi2015optimal,wiezel2016optimization,hatton2017kinematic,bettiol2017purcell} and other locomoting bodies \cite{KaMaRoMe2005a,JiAl2013,yona2019wheeled}. With only three links (and thus only two internal degrees of freedom, the interlink angles), it is easier to consider the full range of possible motions. The low-dimensional configuration space also facilitates optimization studies, by limiting the space of possible motions, and therefore perhaps the number of local minima in the optimized quantity (typically efficiency). Three links are enough to approximate perhaps the most common swimming and crawling motions: undulatory traveling-wave motions \cite{taylor1952analysis,Pu1977a}. With two links, time-periodic motions are limited to reciprocal, scallop-type motions. Here locomotion is possible with fore-aft frictional anisotropy \cite{JiAl2013}, buoyancy \cite{burton2010two}, change of shape \cite{spagnolie2009rehinging}, or when body inertia is considered for sliding bodies \cite{alben2020intermittent}, in which case it may be relatively efficient.

%Energy-optimal strokes for multi-link
%microswimmers: Purcell's loops and Taylor's
%waves reconciled--uses RFT for optimization, optimal strokes centered at origin.

%1-mode, plus 3-mode, plus 5-mode spaces (like Tam and Hosoi)
%(remove omega1--omega2 symmetry--still zero rotation)

\begin{figure} [h]
           \begin{center}
           \begin{tabular}{c}
               \includegraphics[width=6.5in]{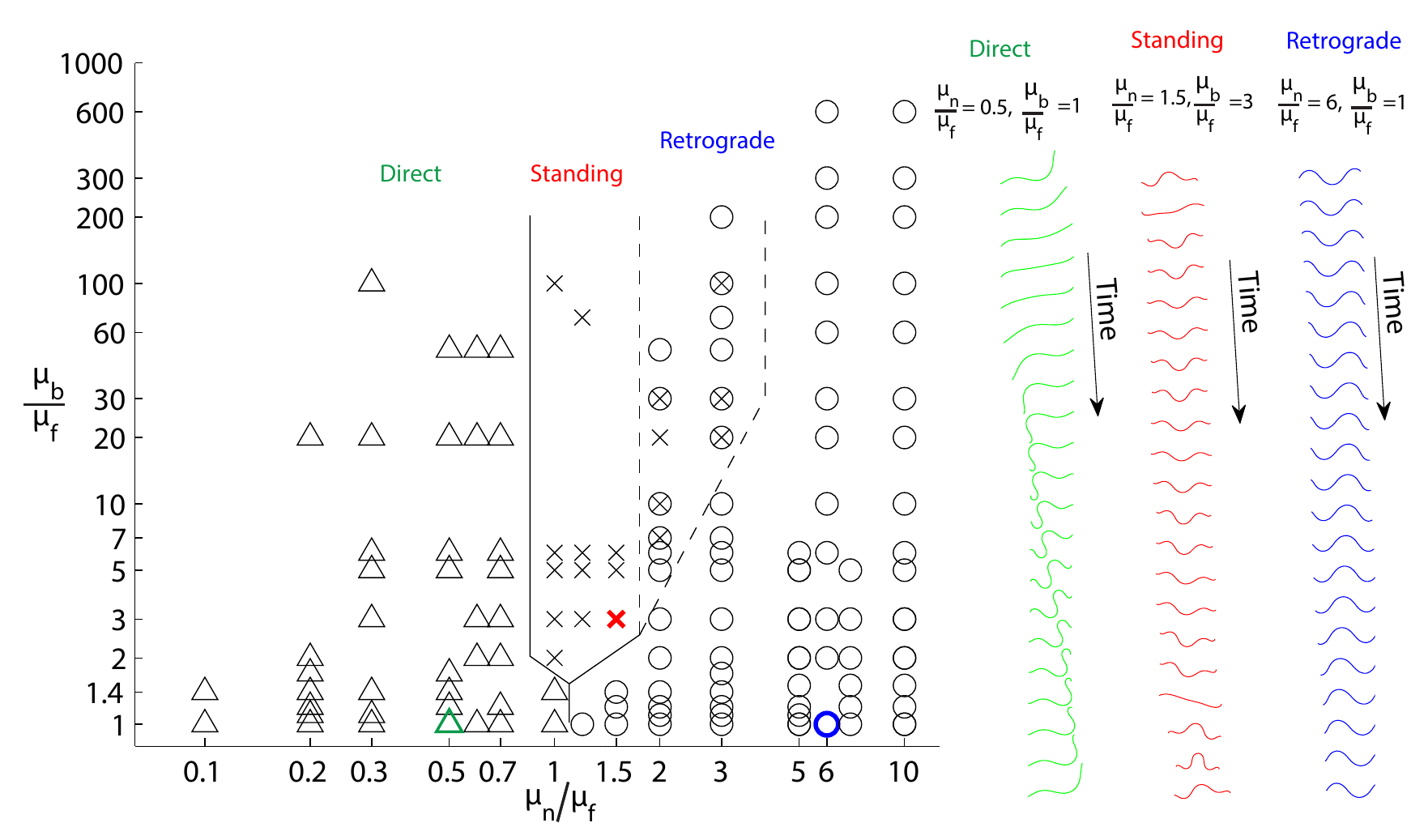} \\
           \vspace{-.25in} \hspace{-.25in}
           \end{tabular}
          \caption{\footnotesize Left: Classification of local optima across friction
coefficient space, presented in \cite{AlbenSnake2013}. Right: Three sequences of snapshots of locally optimal motions giving
examples of direct, standing, and retrograde waves. These occur at particular friction coefficient ratios, listed above the snapshots
and marked with green, red, and blue symbols in the panel at left. The three sequences of snapshots are given
over one period of motion, and displaced vertically to
enhance visibility but with the actual horizontal displacement.
 \label{fig:PhasePlaneMotionsIsotropicAltFig}}
           \end{center}
         \vspace{-.10in}
        \end{figure}

Studies have also considered bodies with more than three links \cite{Chernousko:2003fq,chernousko2005modelling,alouges2013self,alouges2019energy}, obtaining some of the benefits of simplifying the body's spatial configuration while approaching the case of a smooth body. In an earlier work, we computed the optimally efficient sliding motions of a smooth curvilinear body, using a quasi-Newton local optimization algorithm starting from various random initial points in the space of time-periodic body kinematics \cite{AlbenSnake2013}. We truncated the number of shape degrees of freedom at 45 in most cases, superposing products of five spatial modes with
nine temporal modes. We computed optima across a space of friction anisotropy ratios (shown in figure \ref{fig:PhasePlaneMotionsIsotropicAltFig}), i.e. the ratios of friction coefficients for sliding in the normal direction (values on the horizontal axis) and backward direction (values on the vertical axis), relative to the coefficient of friction in the forward direction, which is generally the smallest for real snakes \cite{HuSh2012a}. The model originated in previous experimental and theoretical studies of snake and snake-robot locomotion \cite{ma2001analysis,sato2002serpentine,GuMa2008a,HuNiScSh2009a,HuSh2012a}, analogous to resistive force theory for swimmers \cite{taylor1952analysis,lauga2009hydrodynamics}. In the rightmost portion of the parameter space in figure \ref{fig:PhasePlaneMotionsIsotropicAltFig} ($\mu_n/\mu_f \gg 1$), the optimally efficient motions are relatively smooth retrograde traveling wave motions, and are relatively unchanged when the number of spatial and temporal degrees of freedom in the body kinematics are approximately doubled (from five to ten, and nine to nineteen, respectively). In the central and left parts of the parameter space in figure \ref{fig:PhasePlaneMotionsIsotropicAltFig}, some local optima were identified that are standing and direct wave motions, respectively. However, in these regions there were many optima that were difficult to classify, and it was difficult to obtain convergence from many of the random initial conditions, and to identify global optima. Therefore, in this work we limit the number of spatial degrees of freedom by considering three-link bodies. One advantage is easier visualization of the trajectories in the space of body shapes, which is two-dimensional. With fewer degrees of freedom, optimization is also easier, and we can more completely describe local optima throughout friction coefficient space. Another advantage is that we can go beyond optimization and describe the entire space of possible kinematics to some degree, not just the kinematics that are optimally efficient. At the end of the paper, we employ a stochastic optimization algorithm, which has some robustness advantages over that in \cite{AlbenSnake2013}, to compute optimal three-link kinematics with many temporal modes. We also use it to compute optimal three-link motions with a linear resistance law, corresponding to swimming in or sliding on a viscous medium, and compare with the optima for dry-friction resistance. In \cite{JiAl2013}, we computed optimal kinematics of three-link bodies with up to two frequencies, at a particular choice of friction coefficient ratios motivated by the experiments in \cite{HuNiScSh2009a}. Fast computations of locomotion without inertia are facilitated by precomputing ``velocity maps," maps from shape change to displacements and rotations in physical space \cite{KaMaRoMe2005a,hatton2013geometric,JiAl2013}. In \cite{HaCh2011a,hatton2013geometric,dai2016geometric},
velocity maps were used to predict swimming motions that give large net displacements with zero net rotation. In \cite{alben2019efficient}, we developed the iterative method for computing velocity maps with Coulomb friction resistance that is used here, and computed optimal motions of three-link bodies with isotropic friction and a single frequency. Now, we develop a stochastic optimization algorithm that allows us to compute optimal kinematics with many frequencies (up to nine are used here), in a large portion of the two-dimensional space of friction coefficient ratios. We also describe properties of the full space of kinematics, both optimal and nonoptimal.

\section{Model}

\begin{figure} [h]
           \begin{center}
           \begin{tabular}{c}
               \includegraphics[width=6.3in]{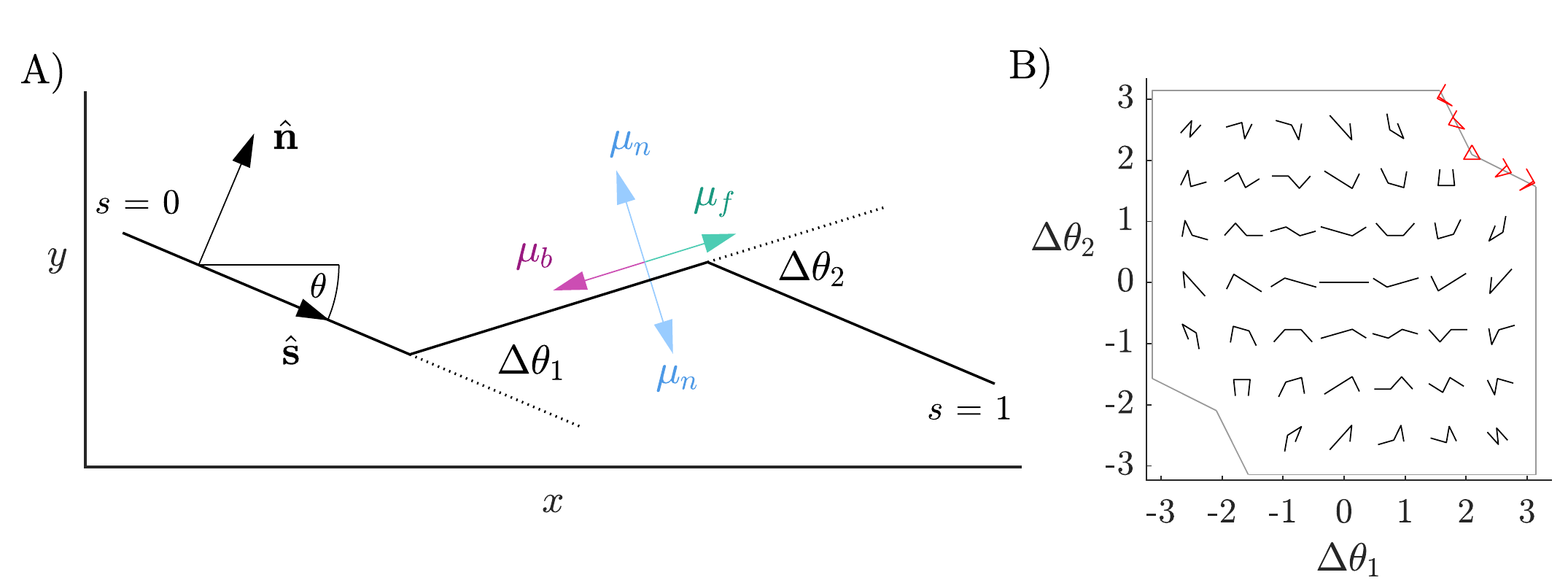} \\
           \vspace{-.25in}
           \end{tabular}
          \caption{\footnotesize A) Schematic diagram of a 
three-link body with changes in angles $\Delta\theta_1$ (here positive) and
$\Delta\theta_2$ (here negative) between the links. The body is
parametrized by arc length $s$ (nondimensionalized by body length),
 at an instant in time.  The tangent angle
and the unit vectors
tangent and normal to the curve at a point are labeled. Vectors representing
forward, backward, and normal velocities are shown with
the corresponding friction coefficients $\mu_f$, $\mu_b$, and $\mu_n$.
B) Examples of body shapes in the ($\Delta\theta_1$, $\Delta\theta_2$)-plane.
Shapes that do not self-intersect are shown in black; a few shapes at the threshold of
self-intersection are shown in red.
 \label{fig:ThreeLinkSchematic}}
           \end{center}
         \vspace{-.10in}
        \end{figure}

We use the same Coulomb-friction
model as \cite{HuNiScSh2009a,HuSh2012a,JiAl2013} and other
recent studies. 
The body is thin compared to its length, so for simplicity 
we approximate its motion
by that of a polygonal curve $\mathbf{X}(s,t) = (x(s,t), y(s,t))$,
parametrized by arc length $s$ and varying with time $t$. A schematic
diagram is shown in figure \ref{fig:ThreeLinkSchematic}A.

The basic problem is to prescribe the time-dependent
shape of the body in order to obtain efficient locomotion. The shape is
described by $\Delta\theta_1(t)$ and $\Delta\theta_2(t)$, the differences
between the tangent angles of the adjacent links. 
A set of possible body shapes are plotted at the corresponding ($\Delta\theta_1$, $\Delta\theta_2$) locations
in figure \ref{fig:ThreeLinkSchematic}B. The region inside the gray polygonal boundary consists of shapes that do not self-intersect. Five examples of shapes that lie on the boundary are shown in red (along the upper right portion of the boundary). In this work we will consider time-periodic kinematics, which are represented by closed curves in the
($\Delta\theta_1$, $\Delta\theta_2$)-plane.

To write the dynamical equations (Newton's laws), we first write
the body tangent angle as $\theta(s,t)$; it satisfies 
$\partial_s x = \cos\theta$ and $\partial_s y = \sin\theta$.
The unit vectors
tangent and normal to the body are $\hat{\mathbf{s}} = (\partial_s x, \partial_s y)$ 
and $\hat{\mathbf{n}} = (-\partial_s y, \partial_s x)$
respectively. We write 
\begin{align}
\theta(s,t) &= \theta_0(t) + \Delta \theta_1(t) H(s-1/3) + \Delta \theta_2(t) H(s-2/3) \label{theta0}
\end{align}
\nn where $H$ is the Heaviside function and $\theta_0(t)$ is the tangent
angle at the ``tail" (the $s=0$ end), an unknown to be solved for using Newton's equations of motion.
%We prescribe the body shape as $\Theta(s,t)$, the tangent angle in the ``body frame,'' 
%defined as a frame that rotates and translates so that at every time the body tail ($s=0$) lies at
%the origin in the body frame and the body has zero tangent angle at the tail ($\Theta(0,t) = 0$).
The body position is obtained by integrating $\theta$:
\begin{align}
x(s,t) & = x_0(t) + \int_0^s \cos \theta(s',t) ds', \label{x0}\\
y(s,t) &= y_0(t) + \int_0^s \sin \theta(s',t) ds'. \label{y0}
\end{align}
\nn The tail position $\mathbf{X}_0(t) = (x_0(t), y_0(t))$ and tangent angle $\theta_0(t)$
are determined by the force and torque
balance for the body, i.e. Newton's second law:
\begin{align}
\int_0^L \rho \partial_{tt} x ds &= \int_0^L f_x ds, \label{fx0} \\
\int_0^L \rho \partial_{tt} y ds &= \int_0^L f_y ds, \label{fy0} \\
\int_0^L  \rho \mathbf{X}^\perp \cdot \partial_{tt} \mathbf{X} ds
&= \int_0^L \mathbf{X}^\perp \cdot \mathbf{f} ds. \label{torque0}
\end{align}
\nn Here  $L$ is the body length, $\rho$ is the body's mass per unit length, and
$\mathbf{X}^\perp = (-y,x)$. For simplicity, the body is assumed to be locally inextensible
so $L$ is constant in time.
$\mathbf{f}$ is the force per unit length on the body
due to Coulomb friction with the ground:
\begin{align}
\mathbf{f}(s,t) &\equiv -\mu_n
\left( \widehat{\partial_t{\mathbf{X}}}_\delta\cdot \hat{\mathbf{n}} \right)\hat{\mathbf{n}}
- \left( \mu_f H(\widehat{\partial_t{\mathbf{X}}}_\delta\cdot \hat{\mathbf{s}})
+ \mu_b (1-H(\widehat{\partial_t{\mathbf{X}}}_\delta\cdot \hat{\mathbf{s}}))\right)
\left( \widehat{\partial_t{\mathbf{X}}}_\delta\cdot \hat{\mathbf{s}} \right)\hat{\mathbf{s}}, \label{frictiondelta} \\ 
\widehat{\partial_t{\mathbf{X}}}_\delta &\equiv \frac{\left(\partial_t x, \partial_t y\right)}{\sqrt{\partial_t x^2 +\partial_t y^2 + \delta^2}}. \label{delta}
\end{align}
\nn 
Again $H$ is the Heaviside function, and $\widehat{\partial_t{\mathbf{X}}}_\delta$ is the normalized velocity,
regularized with a small parameter $\delta = 10^{-3}$ here. Nonzero $\delta$ avoids nonsolvability of the equations in a small number of cases where static friction comes into play, as detailed in \cite{alben2019efficient} in the isotropic case. We find empirically that there is little change in the results (less than 1\% in relative magnitude) when $\delta$ is decreased below $10^{-3}$.

According to (\ref{frictiondelta}) the body experiences
friction with different coefficients for motions in different directions with respect to the body.
The frictional coefficients are $\mu_f$, $\mu_b$, and $\mu_n$ for motions
in the forward ($\hat{\mathbf{s}}$), backward ($-\hat{\mathbf{s}}$),
and normal ($\pm\hat{\mathbf{n}}$) directions, respectively. 
If $\mu_b \neq \mu_f$, we define the forward direction so that
$\mu_f < \mu_b$, without loss of generality.  
In general the body velocity at a given point has both tangential and
normal components, and the frictional force density has
components acting in each direction. A similar decomposition of force
into directional components
occurs for viscous fluid forces on slender bodies \cite{cox1970motion}. 

We assume that the body shape $(\Delta\theta_1(t), \Delta\theta_2(t))$ is periodic in time with period $T$,
as is typical for steady locomotion \cite{HuNiScSh2009a}.
We nondimensionalize equations (\ref{fx0})--(\ref{torque0}) by dividing
lengths by the body length $L$, time by $T$, and mass by $\rho L$. Dividing
both sides by $g$ we obtain:
\begin{align}
\frac{L}{gT^2} \int_0^1 \partial_{tt} x ds &= \int_0^1 f_x ds, \label{fxa} \\
\frac{L}{gT^2}\int_0^1 \partial_{tt} y ds &= \int_0^1 f_y ds, \label{fya} \\
\frac{L}{gT^2}\int_0^1 \mathbf{X}^\perp \cdot \partial_{tt} \mathbf{X} ds
&= \int_0^1 \mathbf{X}^\perp \cdot \mathbf{f} ds. \label{torquea}
\end{align}
\nn In (\ref{fxa})--(\ref{torquea}) and from now on, all variables are
dimensionless. If the body accelerations are not very large, as is often the case
for robotic and real snakes \cite{HuNiScSh2009a}, $L/gT^2 \ll 1$,
which means that the body's inertia is negligible. By setting
inertia---and the left hand sides of (\ref{fxa})--(\ref{torquea})---to zero, we simplify the equations
considerably:
\begin{align}
\int_0^1 f_x ds &= \int_0^1 f_y ds = \int_0^1 \mathbf{X}^\perp \cdot \mathbf{f} ds = 0. \label{ftb}
\end{align}
\nn  Similar models were used in 
\cite{GuMa2008a,HuNiScSh2009a,HuSh2012a,JiAl2013,AlbenSnake2013,wang2014optimizing,wang2018dynamics},
and the same model was found to agree well with the motions of biological
snakes in \cite{HuNiScSh2009a}. 

The distance traveled by the body's center of mass over one period is
\begin{align}
d = \sqrt{ \left( \int_0^1 x(s,1) - x(s,0) ds\right)^2 +
\left(\int_0^1 y(s,1) - y(s,0) ds\right)^2}, \label{dist}
\end{align}
\nn also equal to the time-averaged speed of the center of mass, $\| \overline{\partial_t \mathbf{X}} \|$, where the overbar denotes time- and space- ($t$- and $s$-) average.
The work done by the body against friction over one period is
\begin{align}
W &= \int_0^1 \int_0^1 -\mathbf{f}(s,t) \cdot \partial_t\mathbf{X}(s,t)\, ds\, dt, \label{W}
\end{align}
\nn also equal to the time-averaged power expended against frictional forces, $\langle P \rangle$. As in previous works  \cite{HuNiScSh2009a,HuSh2012a,JiAl2013,AlbenSnake2013}, we define the efficiency of locomotion as 
\begin{align}
\lambda = \frac{d}{W} = \frac{\| \overline{\partial_t \mathbf{X}} \|}{\langle P \rangle}. \label{lambda}
\end{align}
\nn The upper bound on efficiency is
\begin{align}
\lambda_{ub} = \frac{1}{\mbox{min}(\mu_f, \mu_b, \mu_n)}, \label{lambdaub}
\end{align}
\nn corresponding to uniform motion in the direction of least friction, and can be approached by a sequence of particular concertina-like motions, as shown in \cite{alben2019efficient}.
In this work we take the relative efficiency $\lambda/\lambda_{ub}$ as the primary measure of performance. 
For the case of zero body inertia considered here, we explained in \cite{alben2019efficient} that $d$, $W$, $\lambda$,
and the body motion are unchanged (in the limit $\delta \to 0$; to a very good approximation for $\delta = 10^{-3}$) by nondecreasing reparametrizations of time, i.e. if $t$ is replaced by any nondecreasing function $\alpha(t)$ that maps the unit interval to itself. The locomotor performance only depends on the path traced by the kinematics in the $(\Delta\theta_1, \Delta\theta_2)$-plane. 

\section{Single-frequency (elliptical) kinematics}

\begin{figure} [h]
           \begin{center}
           \begin{tabular}{c}
               \includegraphics[width=6.5in]{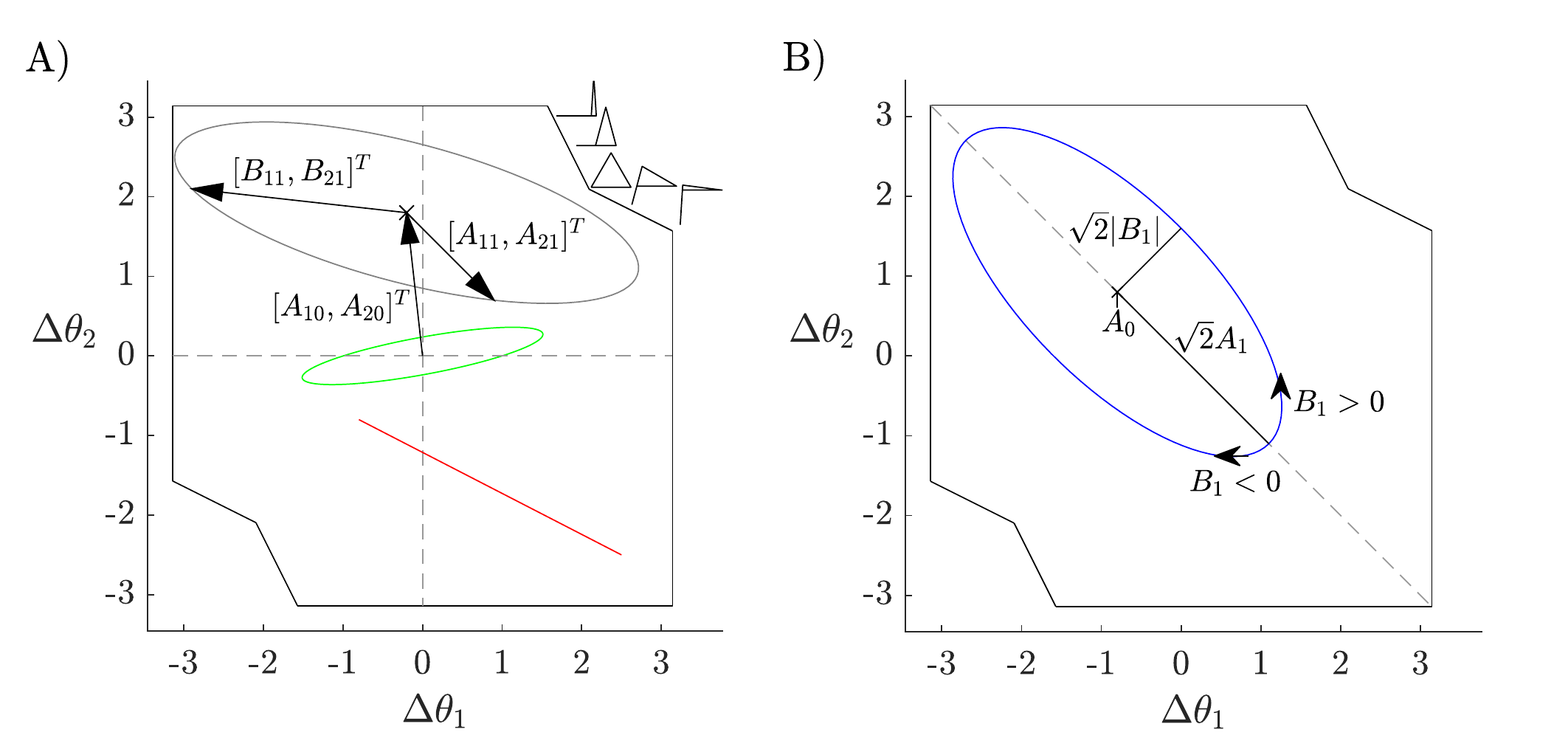} \\
           \vspace{-.25in} \hspace{-.25in}
           \end{tabular}
          \caption{\footnotesize A) Examples of elliptical trajectories in the region of 
non-self-intersecting configurations (inside the black polygonal outline). Examples of body 
configurations at the boundary
of the region are shown at upper right. The gray ellipse has center $A_{10}, A_{20}$ and
shape given by $\{A_{11}, A_{21},B_{11}, B_{21}\}$. We take
$A_{21}=-A_{11}$, which fixes an arbitrary phase.
B) ($\Delta\theta_1(t)$, $\Delta\theta_2(t)$) for a three-link body, symmetric about the line
$\Delta\theta_1 = -\Delta\theta_2$. $A_0$ is the average of $\Delta\theta_1$ over
the ellipse and $\sqrt{2}A_1$ and $\sqrt{2}|B_1|$ are the semi-major and
semi-minor axes of the ellipse. The sign of $B_1$ gives the direction in which the
path is traversed. 
 \label{fig:PathsSchematicEllipses}}
           \end{center}
         \vspace{-.10in}
        \end{figure}

We begin by considering body kinematics that have a single frequency, given by elliptical trajectories in the 
($\Delta\theta_1$, $\Delta\theta_2$)-plane: 
\begin{align}
\Delta\theta_1(t) = A_{10} + A_{11}\cos(2\pi t) + B_{11}\sin(2\pi t), \quad
\Delta\theta_2(t) = A_{20} + A_{21}\cos(2\pi t) + B_{21}\sin(2\pi t), \quad 0 \leq t \leq 1. \label{GenEllipse}
\end{align}
An example is the gray ellipse in figure \ref{fig:PathsSchematicEllipses}A, with the coefficient values shown as vectors.
For any path (\ref{GenEllipse}), the path is unchanged when $t$ is shifted by an arbitrary constant. This arbitrary phase may be fixed in various ways; here we take $A_{21} = -A_{11} \leq 0$, so that at $t=0$ the body shape is at the point on the ellipse located
at angle $-\pi/4$ with respect to the ellipse center. The space of elliptical paths is therefore five-dimensional. 

As in previous works \cite{TaHo2007a,alouges2019energy}, we pay particular attention to the subset of paths that yield no net rotation of the body over one cycle, because these are the kinematics that yield nonzero net locomotion over a long-time average. If there is a nonzero net rotation, points on the body move along circles over large times, so the long-time average velocity is zero. However, such kinematics could still yield efficient locomotion over short-to-medium times, particularly if the net rotation is small. We consider this possibility later.  In \cite{alben2019efficient} we showed that no net rotation occurs for
paths that have a certain bilateral symmetry, under reflection in the line $\Delta\theta_1 = -\Delta\theta_2$, e.g. the blue ellipse in panel B. In that work we discussed the case $\mu_b = \mu_f$, but the same argument holds if $\mu_b \neq \mu_f$. The rotation that occurs as the body traverses the half-ellipse above the line $\Delta\theta_1 = -\Delta\theta_2$ is cancelled by the rotation that occurs on the half-ellipse below the line. Ellipses with bilateral symmetry can
be parametrized as
\begin{align}
\Delta\theta_1(t) = A_{0} + A_{1}\cos(2\pi t) + B_{1}\sin(2\pi t), \quad
\Delta\theta_2(t) = -A_{0} - A_{1}\cos(2\pi t) + B_{1}\sin(2\pi t), \quad 0 \leq t \leq 1. \label{BilatEllipse}
\end{align}
\nn with only three parameters versus five for general ellipses. We may take $A_1 \geq 0$ without loss of generality, by shifting $t \to t +1/2$ if necessary, which leaves the path unchanged.

Another set of paths that yield no net rotation are those with antipodal symmetry, i.e. symmetry with respect to reflection in the origin, such as the green ellipse in panel A. At antipodal points, $\Delta\theta_1$ and $\Delta\theta_2$ are reversed in sign, and so are $\partial_t\Delta\theta_1$ and $\partial_t\Delta\theta_2$. Therefore, the shapes and kinematics of the body are mirror images when viewed in the body frame---defined here as the frame in which the tail lies at the origin, with zero tangent angle. The equations (\ref{ftb}) are solved by equal and opposite values of $d\theta_0(t)/dt$ and mirror image
vectors $d\mathbf{X}_0/dt$ in the body frame, because they result in mirror-image distributions of $\mathbf{f}$ in the body frame, which both satisfy equations (\ref{ftb}). Hence the body rotations at antipodal points cancel, and the net rotation over a full path is zero. Ellipses with antipodal symmetry are also parametrized by three parameters
\begin{align}
\Delta\theta_1(t) = A_{11}\cos(2\pi t) + B_{11}\sin(2\pi t), \quad
\Delta\theta_2(t) =  -A_{11}\cos(2\pi t) + B_{21}\sin(2\pi t), \quad 0 \leq t \leq 1. \label{AntipEllipse}
\end{align}
\nn where $A_{21}$ has again been set to $-A_{11}$ to fix the arbitrary phase.

A third special case that we discuss later is reciprocal kinematics---kinematics that are the same under time reversal. These are degenerate ellipses that reduce to straight line segments, e.g. the red line in panel A. These yield no net locomotion if $\mu_b = \mu_f$ but can yield efficient locomotion in other cases.

\subsection{Efficient single-frequency kinematics}

\begin{figure} [h]
           \begin{center}
           \begin{tabular}{c}
               \includegraphics[width=6.5in]{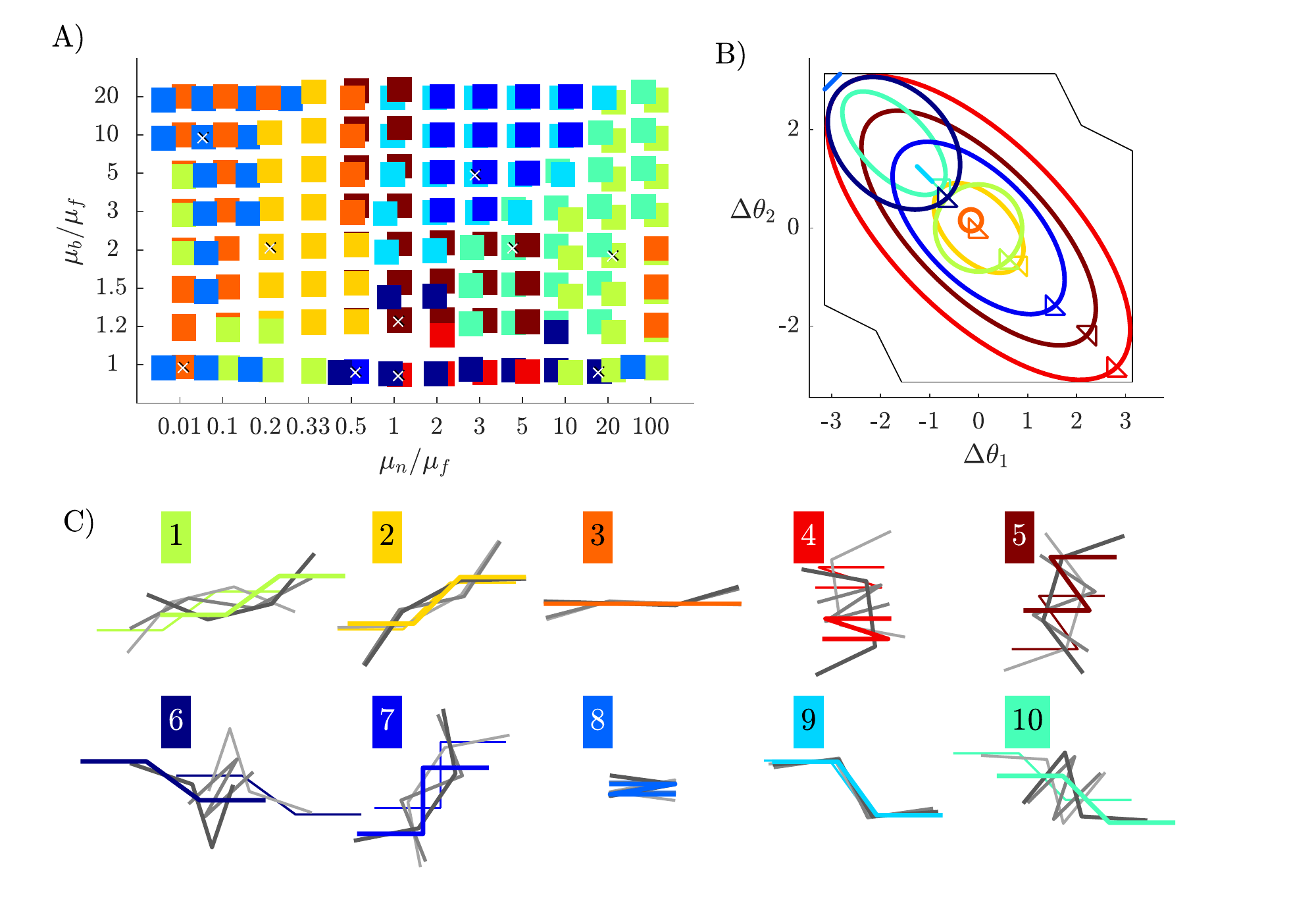} \\
           \vspace{-.25in} \hspace{-.25in}
           \end{tabular}
          \caption{\footnotesize A) Cluster classification of the top two local optima in efficiency for elliptical trajectories across a grid of ($\mu_n/\mu_f$, $\mu_b/\mu_f$) values. The set of 192 local optima, plotted as colored squares, are used to define 10 clusters based on proximity in ($A_0$, $A_1$, $B_1$) space. Near each 
($\mu_n/\mu_f$, $\mu_b/\mu_f$) pair, the top two local optima are represented by two offset but partly overlapping squares. The color of each square denotes the cluster to which it belongs. B) The elliptical trajectory of the optimum closest to the centroid of each cluster, with color corresponding to that cluster. The location of each ellipse is marked by black and white crosses in panel A (for visibility on light and dark backgrounds), at the center of the corresponding square. C) For each ellipse in B, snapshots of the body motion at five instants spaced 1/4-period apart, starting from the thin colored line, proceeding from light to dark gray, and ending with the thick colored line.
 \label{fig:Mode1SymmOptimaMotions}}
           \end{center}
         \vspace{-.10in}
        \end{figure}

We begin by studying the performance of trajectories given by ellipses with bilateral symmetry (e.g. figure \ref{fig:PathsSchematicEllipses}B). We consider ($A_0$, $A_1$, $B_1$) ranging over a three-dimensional grid
in which $A_0$ and $B_1$ range from $-1.2\pi$ to 1.2$\pi$, and $A_1$ from 0 to 1.2$\pi$, each in increments of $\pi/20$. Outside these coefficient ranges, elliptical trajectories are generally not valid because they contain self-intersecting body shapes. We thus obtain a set that fills the space of kinematically-valid ellipses somewhat densely. For the ellipses that lie entirely in the non-self-intersecting region (about 8000), we compute the body motions, work done against friction, and the relative efficiency $\lambda/\lambda_{ub}$ using precomputed velocity maps, as described in \cite{alben2019efficient}. We compute the results for the friction coefficient ratios ($\mu_n/\mu_f$, $\mu_b/\mu_f$) ranging over a 12-by-8 grid with values ranging widely in magnitude, shown on the axes of figure \ref{fig:Mode1SymmOptimaMotions}A. For each ($\mu_n/\mu_f$, $\mu_b/\mu_f$) pair, we compute the top two local optima for efficiency, obtaining $12\times 8 \times 2$ = 192 optima in total. We then use a k-means clustering algorithm (the kmeans function in Matlab) to partition the optima into 10 clusters based on location in ($A_0$, $A_1$, $B_1$)-space.
In figure \ref{fig:Mode1SymmOptimaMotions}A, two colored squares are plotted near each ($\mu_n/\mu_f$, $\mu_b/\mu_f$) pair, the colors denoting the cluster to which each of the two optima at that ($\mu_n/\mu_f$, $\mu_b/\mu_f$) pair belong. The squares with different colors are displaced slightly so both can be seen at a given ($\mu_n/\mu_f$, $\mu_b/\mu_f$) pair. Panel B shows trajectories for the optimum closest to the centroid of each cluster (locations labeled by black and white crosses in panel A). Panel C shows snapshots of the body motions corresponding to each of the 10 ellipses in B. Each sequence of snapshots starts from the thin colored line, proceeds from light gray to dark gray, and ends with the thick colored line. An animation of these motions is shown in the supplemental material.

We see in panel A that each cluster (i.e. color) tends to occur in certain contiguous regions of ($\mu_n/\mu_f$, $\mu_b/\mu_f$)-space. In other words, the friction coefficient ratios tend to select certain types of motions as optima. In figure \ref{fig:PhasePlaneMotionsIsotropicAltFig} (from results in \cite{AlbenSnake2013}), we sorted the optima for smooth bodies into three wave-like motions. It was difficult to obtain convergence to local optima at many ($\mu_n/\mu_f$, $\mu_b/\mu_f$) values in the smooth case. Also, many of the optima in \cite{AlbenSnake2013} were difficult to classify, and did not correspond to the wave-like classification. With the smaller parameter space represented by elliptical trajectories of three-link bodies, here we are able to identify all local optima, and sort them more precisely. Unlike the three wave-type categories, the ten clusters here cover all of kinematic parameter space (given by ($A_0$, $A_1$, $B_1$)). In panel A, the ten clusters overlap in multiple ways, but seven major regions in ($\mu_n/\mu_f$, $\mu_b/\mu_f$)-space can be identified: 
\begin{enumerate}
\item $\mu_n/\mu_f \ll 1$, represented by optima 1, 3, and 8 (as numbered in panel C). Optima 3 and 8 have very small amplitudes about motions that are nearly flat or completely folded, respectively, and move with a slight motion mainly in the normal direction when $\mu_n/\mu_f \ll 1$. 
\item $0.1 < \mu_n/\mu_f < 1$, represented by optimum 2. This is a somewhat larger amplitude version of optimum 3, and translates in both normal and tangential directions.
\item In the vicinity of isotropic friction, $\mu_n/\mu_f \approx 1$, $\mu_b/\mu_f \approx 1$, is a heterogeneous region in which two large-amplitude motions (4 and 6) predominate.
\item $\mu_n/\mu_f \approx 1$, $\mu_b/\mu_f > 1$. The brown optimum (5) is the most common here. It is a large-amplitude motion that translates roughly tangent to the body's mean flat state. This is a heterogeneous region with both small and large-amplitude motions (3, 6, and 9); 
\item $\mu_n/\mu_f > 1$, $\mu_b/\mu_f > 1$ but not $\gg 1$. The optima are mainly 5 and 10, both large-amplitude motions; 
\item $\mu_n/\mu_f > 1$, $\mu_b/\mu_f \gg 1$. Here the optima are mainly 7 (a large amplitude motion) and 9 (a very small amplitude motion); 
\item  $\mu_n/\mu_f \gg 1$. Here 1 and 10 predominate, and the body moves mainly in the tangential direction. 10 roughly resembles concertina motion of snakes, in which the front and rear of the body contract and expand alternately, while 1 resembles lateral undulation, i.e. a traveling wave along the body. 
\end{enumerate}

Like the smooth case, the three-link case shows a rough partition based on small, medium, and large values of $\mu_n/\mu_f$, with additional divisions based on $\mu_b/\mu_f$. It is interesting that at most ($\mu_n/\mu_f$, $\mu_b/\mu_f$) values, the top two optima come from two different clusters. One might have expected the top two optima to be nearby motions within the same cluster. This is the case in most of the region where yellow squares are found, but is rarely true elsewhere. This could result from a relatively smooth efficiency landscape in most cases, without large numbers of closely spaced optima. Six of the ten optima in panel B are symmetric or nearly symmetric about the origin, meaning they oscillate about a flat mean shape. The remaining four (6, 8, 9, and 10) oscillate about mean shapes that are folded to a large extent. We also find that the undulatory optimum 1 is common both at $\mu_n/\mu_f \ll 1$ and $\gg 1$, but not at intermediate values (similar kinematics give zero net locomotion with isotropic friction \cite{alben2019efficient}). The small amplitude motions 3 and 8 also appear where $\mu_n/\mu_f$ is very small and very large. 

\begin{figure} [h]
           \begin{center}
           \begin{tabular}{c}
               \includegraphics[width=6in]{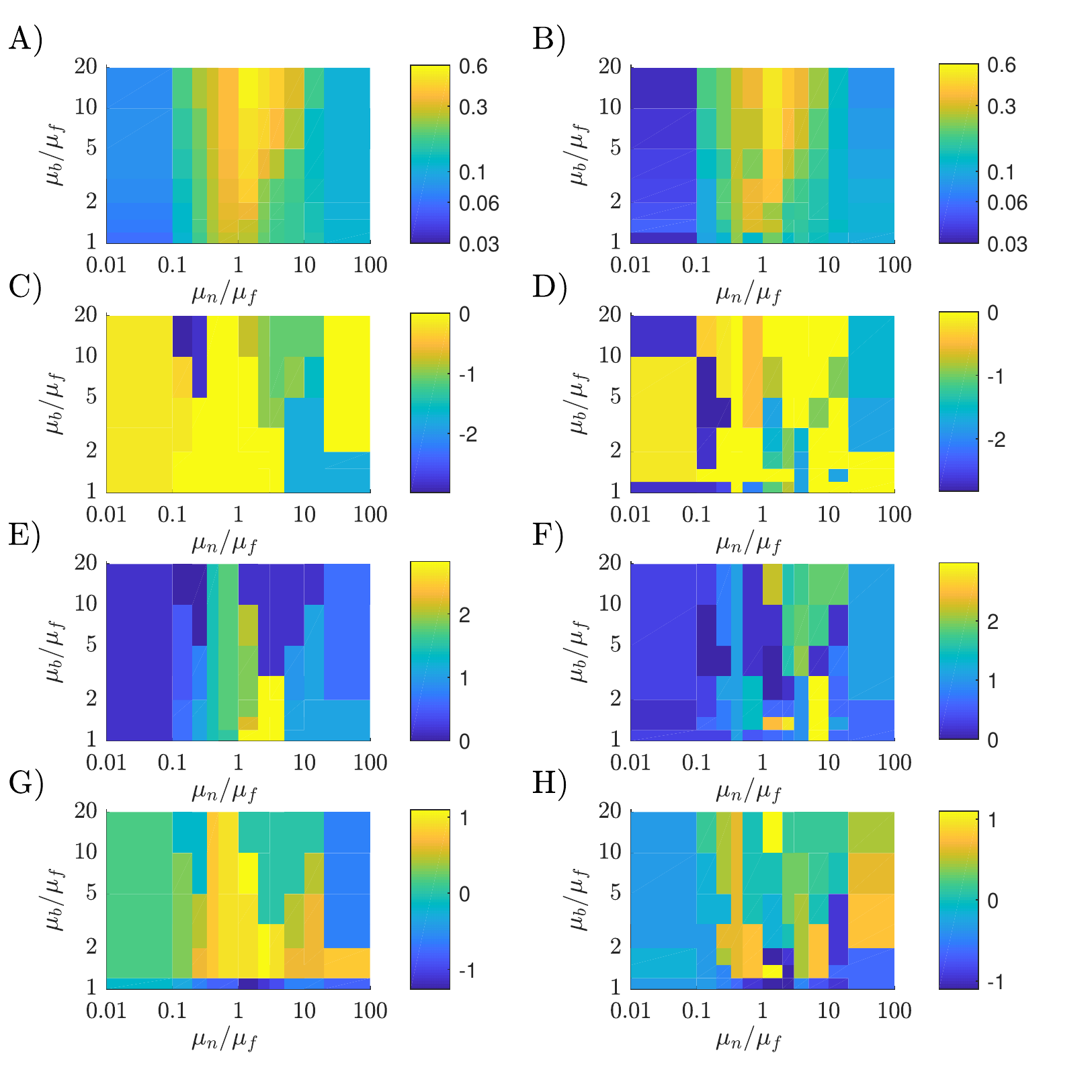} \\
           \vspace{-.25in} \hspace{-.25in}
           \end{tabular}
          \caption{\footnotesize Relative efficiencies of the global (A) and second-best optima (B) among elliptical trajectories with bilateral symmetry, and the corresponding kinematic parameters $A_0$ (C and D, respectively),
$A_1$ (E and F), and $B_1$ (G and H). 
 \label{fig:Mode1SymmOptimaParams}}
           \end{center}
         \vspace{-.10in}
        \end{figure}

Figure \ref{fig:Mode1SymmOptimaParams} shows the relative efficiencies of the global (panel A) and second-best optima (panel B) in these regions. The corresponding $A_0$, $A_1$, and $B_1$ are plotted in panels C, E, and G for the global optima and in panels D, F, and H for the second best optima, respectively. The maximum relative efficiency, nearly 0.6, is achieved at $\mu_n/\mu_f = 1$ and $\mu_b/\mu_f = 20$, the top center of panel A, by kinematics in the cluster represented by motion 9 in figure 
\ref{fig:Mode1SymmOptimaMotions}C---a very small amplitude reciprocal kinematics. The second best optimum at the same friction coefficient ratios (top center of panel B), is nearly as good, 
but corresponds to a very different kinematics---number 5 in figure
\ref{fig:Mode1SymmOptimaMotions}C. The motion shown there is for an optimum at the same $\mu_n/\mu_f$ but a much smaller $\mu_b/\mu_f$ (1.2). The maximum relative efficiencies decline smoothly and monotonically in all directions moving away from the top center of panel A. At the bottom center of panel A is isotropic friction, with maximum relative efficiency 0.26. The kinematics are given by the large red ellipse in figure \ref{fig:Mode1SymmOptimaMotions}B and the motion is number 4 in panel C. Moving to the lower left corner of figure \ref{fig:Mode1SymmOptimaParams}A, $\mu_n/\mu_f$ = 0.01 and $\mu_b/\mu_f$ = 1,
the relative efficiency drops to 0.06, its minimum over the panel, given by motion 3 in figure \ref{fig:Mode1SymmOptimaMotions}C. Here, even a small amount of tangential motion causes a large drop in relative efficiency. At the other extreme, $\mu_n/\mu_f$ = 100, 
the relative efficiency is 0.1, and is achieved by a small-amplitude circular trajectory about the origin (the flat state), similar to the kinematics of motion 3 in figure \ref{fig:Mode1SymmOptimaMotions}C, but now resulting in mainly {\it tangential} motion. For both $\mu_n/\mu_f \ll 1$ and $\gg 1$, the single frequency and the three-link body do not permit sufficiently fine scale motions to come close to the upper bound of efficiency. We will see later that adding more frequencies allows a large improvement in efficiency for $\mu_n/\mu_f \ll 1$, but less so for $\mu_n/\mu_f \geq 1$, for three-link bodies.

The relative efficiencies of the second-best optima, shown in panel B, are 70--99\% of those of the best optima over most of the middle parts of the panels, but drop to 30--60\% of the best values at the most extreme values of
$\mu_n/\mu_f$, 0.01 and 100. The values have a general pattern of decrease from a peak at the top center that is similar to panel A, but with a bit less monotonicity. The mean shape is flat for $A_0$ = 0 and more folded as $|A_0|$ increases. Figure \ref{fig:Mode1SymmOptimaParams}C shows the $A_0$ values for the best optima. They are close to 0 (a nearly flat mean shape) in most cases, except for some very folded cases at top, left of center (i.e. motion 8), and at bottom, right of center (i.e. motion 10). The amplitudes of the motions, described by $A_1$ (panel E) and $B_1$ (panel G), are typically close to 0 for $\mu_n/\mu_f \ll 1$, large for $\mu_n/\mu_f \approx 1$, and then very small again (for $\mu_b/\mu_f \gg 1$) or moderately small (for $\mu_b/\mu_f \approx 1$) when $\mu_n/\mu_f \geq 1$. There is more heterogeneity among the parameters for the second best optima (right column, panels D, F, and H). By tracking the parameters of the top several optima (not shown beyond the top two) across friction coefficient parameter space, we have found that there are distinct branches of optima with $A_0$, $A_1$, and $B_1$ values that change gradually
as the friction coefficient ratios are varied. Their ordering by relative efficiency switches at certain friction coefficient ratios. This accounts for some of the sharp changes in the parameters of the first and second columns at certain friction coefficient values, where the best or second-best optima switch from one branch of optimal motions to another.

\begin{figure} [h]
           \begin{center}
           \begin{tabular}{c}
               \includegraphics[width=7.2in]{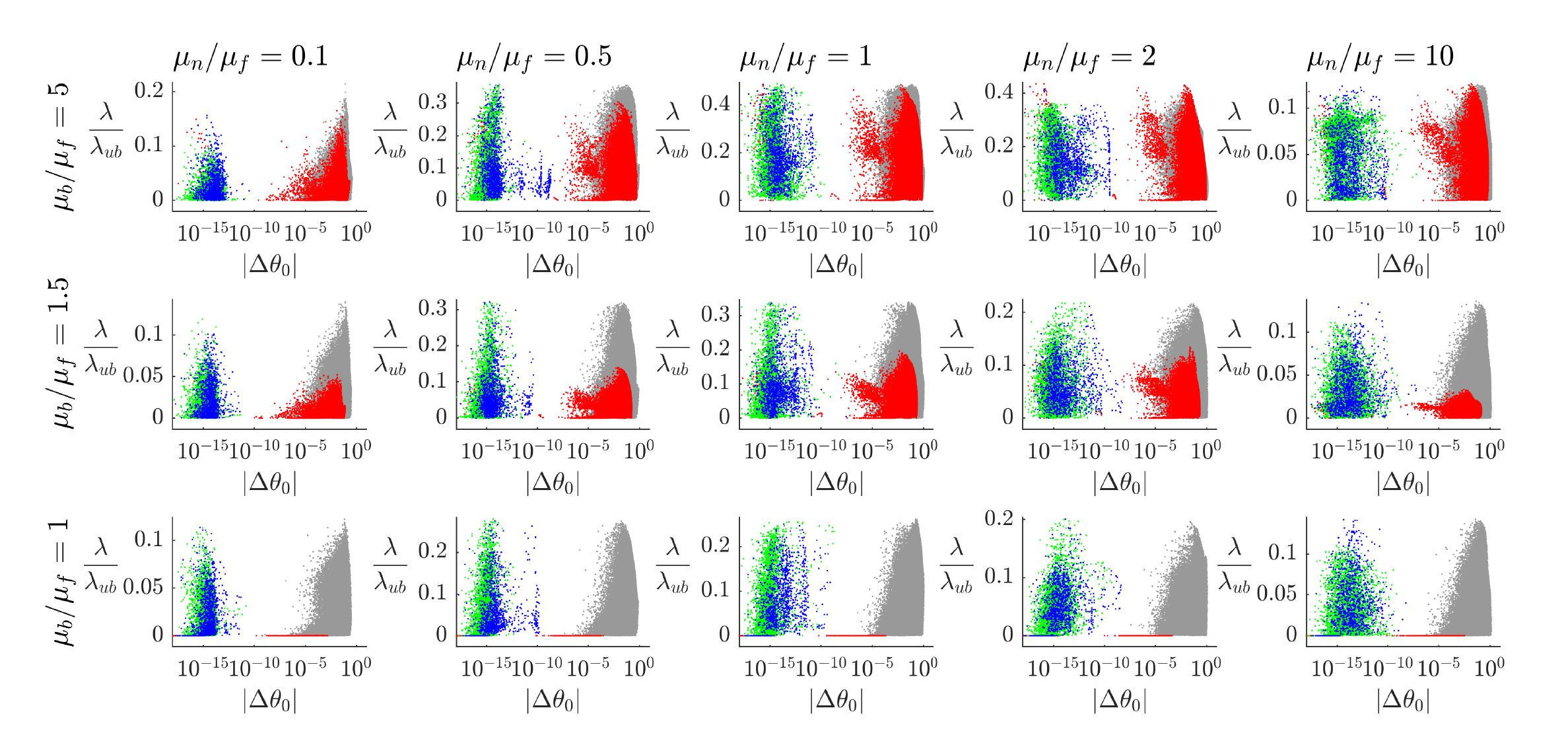} \\
           \vspace{-.25in} \hspace{-.25in}
           \end{tabular}
          \caption{\footnotesize Relative efficiency ($\lambda/\lambda_{ub}$) versus net rotation ($|\Delta \theta_0|$) for elliptical trajectories that are bilaterally symmetric (blue dots), antipodally symmetric (green dots), or reciprocal (red dots). Values for other trajectories are shown by gray dots. Each panel shows data at a given pair of friction coefficient ratios, labeled along the top and left of the figure. 
 \label{fig:RotationDataFigure}}
           \end{center}
         \vspace{-.10in}
        \end{figure}

So far we have considered elliptical trajectories with bilateral symmetry, a three-parameter space. We now
enlarge to the full five-parameter space of arbitrary elliptical trajectories, most of which have nonzero net rotations. We investigate to what extent efficient locomotion can occur with nonzero, but small (possibly very small) net rotation. If some nonsymmetric motions have negligible rotation and greatly outperform the symmetric cases with zero net rotation (exemplified by the green and blue ellipses in figure \ref{fig:PathsSchematicEllipses}), we should consider the larger space of nonsymmetric motions further. We consider the general
ellipse in (\ref{GenEllipse}), fixing the phase by taking $A_{21} = -A_{11}$ and $A_{11} \geq 0$ as discussed previously. Each parameter varies from $-1.2\pi$ to 1.2$\pi$ (except $A_{11}$, varying from 0 to 1.2$\pi$) in increments of $\pi/20$. Restricting to paths in the region of non-self-intersecting bodies, we obtain 4.7 $\times 10^6$ ellipses (compared to about 8000 in the bilaterally symmetric case), a large increase due to exponential growth with parameter space dimension. In figure \ref{fig:RotationDataFigure}, we plot the relative efficiency versus net rotation for the general elliptical trajectories, for various friction coefficient ratios. Each panel has a different set of friction coefficient ratios (labeled at left and top), on a 5-by-3 grid that is a subset of the 12-by-8 grid considered earlier. Each trajectory is represented by a dot, gray for nonsymmetric, blue for bilaterally symmetric, green for antipodally symmetric, and red for reciprocal (as in the examples of figure \ref{fig:PathsSchematicEllipses}). 

The gray dots can have very small rotations, as small as $10^{-8}$ in some cases. However, the green and blue dots are generally orders of magnitude smaller, $\in [10^{-18}, 10^{-10}]$. In most panels,
the green and blue dots achieve top efficiencies that are essentially the same as those of the much larger sets of gray dots.  However, in the first column ($\mu_n/\mu_f = 0.1$), the gray dots reach efficiencies that are 20--30\% higher in each panel. Excluding those with net rotations $> 10^{-2}$ decreases this advantage substantially. Among the gray dots there is a decline in relative efficiency as net rotation tends to zero, and the gray dots with highest efficiencies usually have net rotations $\gtrsim 10^{-3}$. Some of the gray dots are only slight perturbations of symmetric cases, so we would expect similar efficiencies with small but nonzero net rotations. The red dots (reciprocal motions) achieve zero net locomotion, and hence zero relative efficiency, in the bottom row ($\mu_b = \mu_f$). They underperform the other groups in the middle row, but are equal or close to the top performers in the top row, particularly the right side ($\mu_n/\mu_f \geq 1$). In the middle and top rows, most reciprocal motions have nonzero, and sometimes large rotations. However, a small group of red dots can be seen (by zooming in), distinct from the blue and green dots, with very small rotations ($\leq 10^{-15}$), and with high efficiencies. These are nonsymmetric versions of motions 8 and 9 in figure \ref{fig:Mode1SymmOptimaMotions}. Because the green and blue dots achieve nearly the same peak relative efficiencies as the gray dots, and are fewer in number by many orders of magnitude, we consider only these symmetric cases when we add higher frequencies. It rapidly becomes impractical to compute all periodic trajectories with coefficients on the aforementioned grids as the number of coefficients increases above five. Nonsymmetric trajectories with up to two frequencies are described by nine coefficients. Using the same coefficient grids as for the nonsymmetric ellipses (with a single frequency), an estimate of the factor of increase in computing time for the nine-dimensional space relative to the five-dimensional space is $49^4 \approx 6\times 10^6$. Many coefficients lead to self-intersecting trajectories, but even after eliminating these, the factor of increase is many orders of magnitude and beyond our computing resources. Bilaterally symmetric trajectories of a given frequency are described by half the coefficients (plus one) of the nonsymmetric ones, allowing us to consider the full bilaterally symmetric trajectory parameter space with higher frequencies, but only a small number of them. 

%discretize the five parameters  \{$A_{10}, A_{20}, {A_{11}, B_{11}, B_{21}\}$, with 

\section{Multiple-frequency kinematics}

\begin{figure} [h]
           \begin{center}
           \begin{tabular}{c}
               \includegraphics[width=7in]{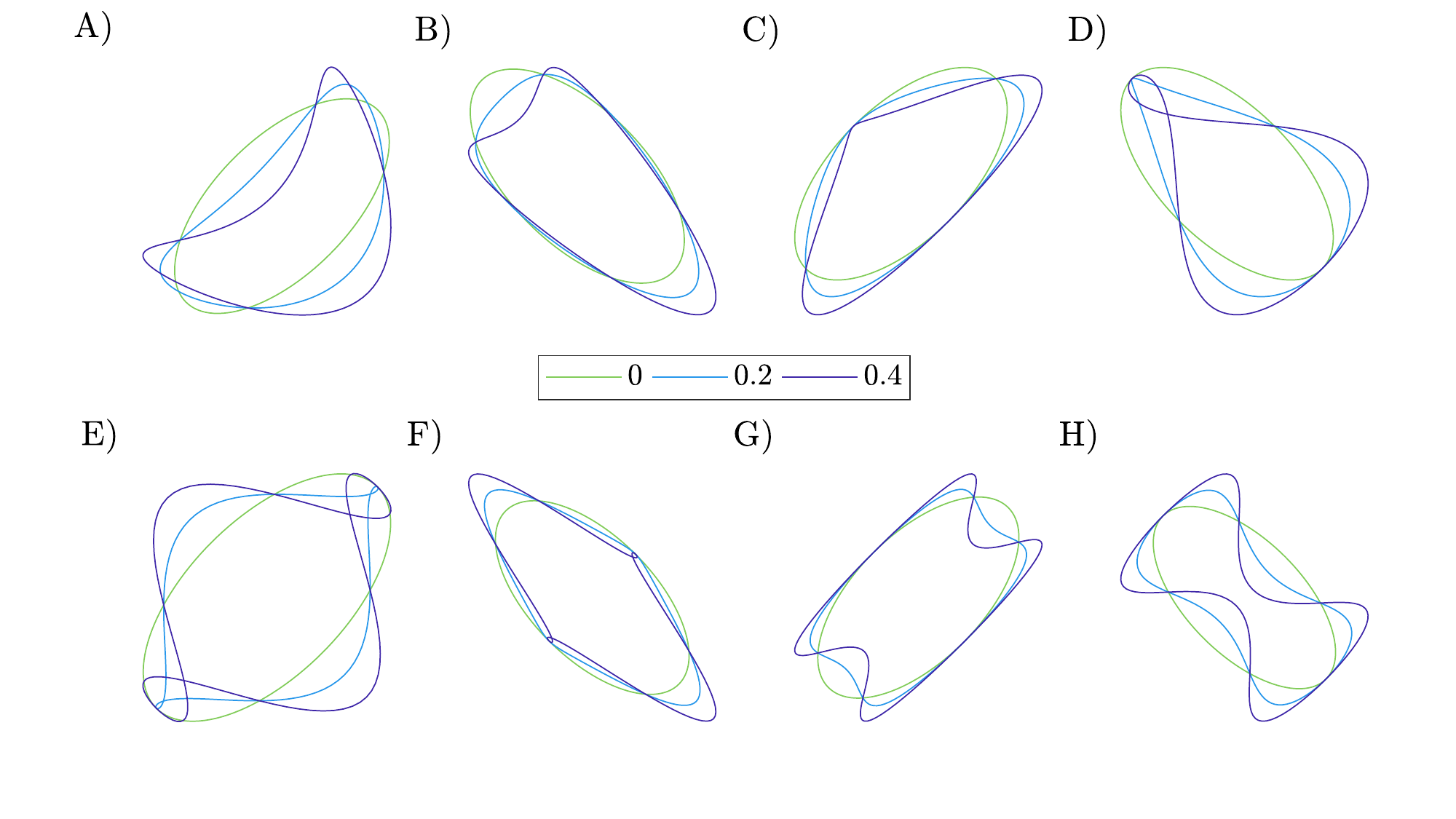} \\
           \vspace{-.25in} \hspace{-.25in}
           \end{tabular}
          \caption{\footnotesize Examples of the effect of adding higher harmonics to elliptical trajectories. The trajectories are given by (\ref{BilatEllipseHigher}). In A, C, E, and G, we have $A_1$ = 0.5 and $B_1$ = 1; in B, D, G, and H, we have $A_1$ = 1 and $B_1$ = 0.5. To these ellipses we add just one additional nonzero mode, setting either $A_2$ (in A and B), $B_2$ in (C and D), $A_3$ (E and F), or $B_3$ (G and H) to 0.2 (light blue lines) or 0.4 (dark blue lines). 
 \label{fig:HigherHarmonicTrajectoriesFig}}
           \end{center}
         \vspace{-.10in}
        \end{figure}

We now add higher frequencies to elliptical trajectories, considering trajectories with bilateral symmetry here (e.g. the blue ellipse in figure \ref{fig:PathsSchematicEllipses}A), and both bilateral and antipodal symmetry later. Trajectories with bilateral symmetry and frequencies up to $k$ are given by 
\begin{align}
\Delta\theta_1(t) = A_0 + \sum_{n = 1}^k A_{n}\cos(2\pi n t) + B_{n}\sin(2\pi n t) \; ; \;
\Delta\theta_2(t) = -A_0 + \sum_{n = 1}^k -A_{n}\cos(2\pi n t) + B_{n}\sin(2\pi n t), \; 0 \leq t \leq 1. \label{BilatEllipseHigher}
\end{align}
while those with antipodal symmetry are given by
\begin{align}
\Delta\theta_1(t) = \sum_{\substack{n = 1\\ n\, \mbox{\scriptsize odd}}}^k A_{1n}\cos(2\pi n t) + B_{1n}\sin(2\pi n t) \; ; \; 
\Delta\theta_2(t) = \sum_{\substack{n = 1\\ n\, \mbox{\scriptsize odd}}}^k A_{2n}\cos(2\pi n t) + B_{2n}\sin(2\pi n t), \; 0 \leq t \leq 1. \label{AntipEllipseHigher}
\end{align}
In both cases we have $2k+1$ terms (when we use $A_{21} = -A_{11}$ to set the arbitrary phase in (\ref{AntipEllipseHigher})) compared to $4k+1$ terms in the general nonsymmetric case, for $k \geq 1$.
%\begin{align}
%\Delta\theta_1(t) &= A_0 + A_{1}\cos(2\pi t) + B_{1}\sin(2\pi t) + A_{2}\cos(4\pi t) + B_{2}\sin(4\pi t) + A_{3}\cos(6\pi t) + B_{3}\sin(6\pi t), \label{BilatEllipseHigher1}\\
%\Delta\theta_2(t) &= -A_0 - A_{1}\cos(2\pi t) + B_{1}\sin(2\pi t) - A_{2}\cos(4\pi t) + B_{2}\sin(4\pi t) - A_{3}\cos(6\pi t) + B_{3}\sin(6\pi t), \quad 0 \leq t \leq 1. \label{BilatEllipseHigher2}
%\end{align}
\nn  
Figure \ref{fig:HigherHarmonicTrajectoriesFig} shows examples of bilaterally symmetric trajectories obtained by adding modes with two or three times the frequency of the basic ellipse. %The trajectories are given by
In both rows, we start with example ellipses shown in green. These have just the $A_1$ and $B_1$ terms in (\ref{BilatEllipseHigher}), with all other terms zero. We take the major axis twice as long as the minor axis in these examples, 
%and have bilateral symmetry about the line $\Delta\theta_1 = -\Delta\theta_2$. 
so in A, C, E, and G, we have $A_1$ = 0.5 and $B_1$ = 1, while in B, D, G, and H, we have the other symmetric orientation, given by $A_1$ = 1 and $B_1$ = 0.5. To these ellipses we add just one additional nonzero mode, setting either $A_2$ (in A and B), $B_2$ in (C and D), $A_3$ (E and F), or $B_3$ (G and H) to 0.2 (light blue lines) or 0.4 (dark blue lines), and the other coefficients to zero. These examples show that the effects of the $4\pi t$ modes (top row) are approximately to dilate the body on one side and contract on the other, though the change of shape is nonuniform and somewhat complicated. The $6\pi t$ modes (bottom row) approximately dilate the body at one pair of opposite sides and contract at the other pair. The trajectories self-intersect in several cases (which is separate from the question of whether the body self-intersects, determined by the location of the trajectory in
($\Delta\theta_1$, $\Delta\theta_2$)-space).  Another, geometric interpretation of the terms in (\ref{BilatEllipseHigher})--(\ref{AntipEllipseHigher}) was given by \cite{mcgarva1993harmonic}: those with lowest frequency (1) represent an ellipse; those with frequency 2 (i.e. with coefficients $A_2$ and $B_2$) also represent an ellipse, but one that is traversed twice within the unit period, and likewise for any frequency $n$. Thus  (\ref{BilatEllipseHigher})--(\ref{AntipEllipseHigher}) can be thought of as superpositions of ellipses which are traversed integer numbers of times within the unit period.

\begin{figure} [h]
           \begin{center}
           \begin{tabular}{c}
               \includegraphics[width=7in]{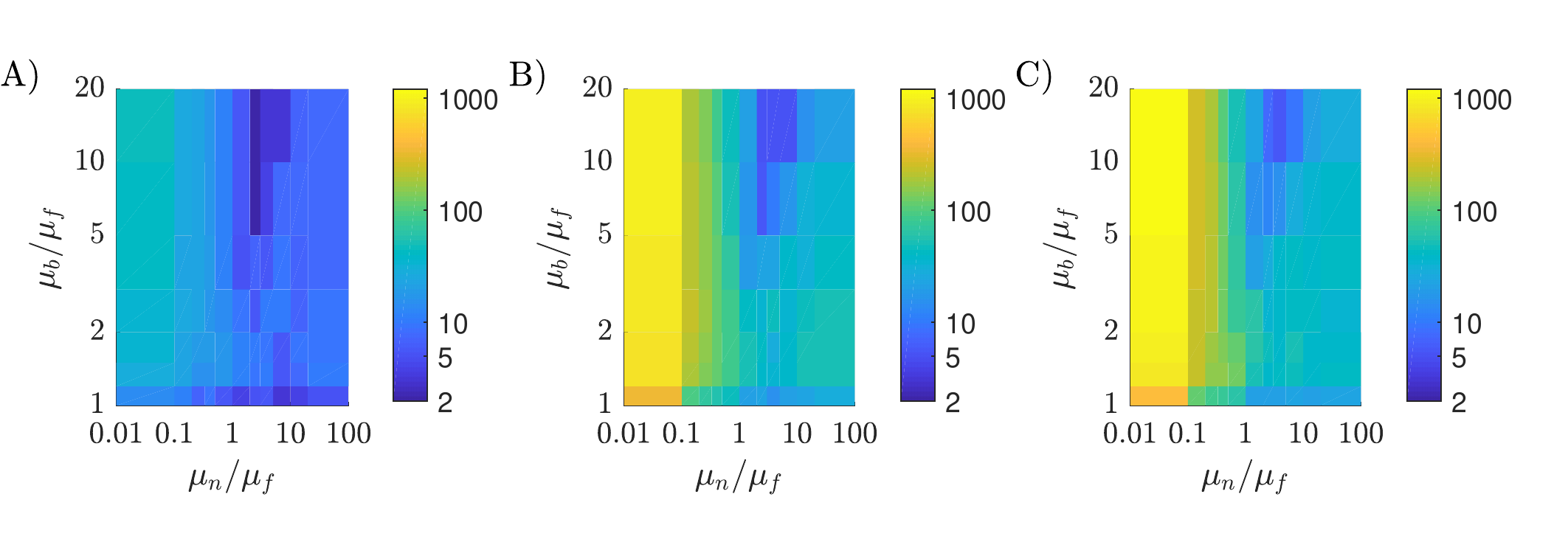} \\
           \vspace{-.25in} \hspace{-.25in}
           \end{tabular}
          \caption{\footnotesize The numbers of local optima of efficiency at various friction coefficient ratios in the space of $\{A_0, A_1, B_1\}$ describing bilaterally symmetric ellipses (panel A), the larger space of $\{A_0, A_1, B_1, A_2, B_2\}$ with second harmonics added (panel B), and the space of $\{A_0, A_1, B_1, A_3, B_3\}$ with third harmonics added (panel C). 
 \label{fig:NumOptimaFig}}
           \end{center}
         \vspace{-.10in}
        \end{figure}

It is very expensive to solve for the body motions for trajectories of the form (\ref{BilatEllipseHigher}) with $k > 2$ with a dense grid of coefficients, i.e. varying all $2k+1$ coefficients on the aforementioned grids with spacing $\pi/20$. Instead, we consider two five-dimensional subspaces, the first consisting of ellipses plus second harmonics, varying $\{A_0, A_1, B_1, A_2, B_2\}$ on the aforementioned grids, and the second consisting of ellipses plus third harmonics, i.e. varying $\{A_0, A_1, B_1, A_3, B_3\}$ on the same grids. In figure \ref{fig:NumOptimaFig} we plot the numbers of local optima for efficiency at various friction coefficient ratios. This quantity gives a measure of the smoothness of efficiency space. The number of optima for bilaterally symmetric ellipses, i.e. the space of $\{A_0, A_1, B_1\}$, are shown in panel A; ellipses plus second harmonics are shown in panel B; and ellipses plus third harmonics are shown in panel C. In panel A, the number of local optima has a minimum of two at the top, right of center, and a maximum of 45 at the top left. These are also locations where the relative efficiency was large and small for the best elliptical trajectories, according to figure \ref{fig:Mode1SymmOptimaParams}A. On the right side of figure \ref{fig:NumOptimaFig}A ($\mu_n/\mu_f > 1$), there are at most 10 optima, and about 2--4 times as many at points with the reciprocal value of $\mu_n/\mu_f$, on the left side.  In panels B and C, the numbers of local optima increase enormously at the top left to about 1000 in each case, while the minimum value of 2 in A increases modestly, to 4 and 6 in B and C, respectively. At other points, the numbers of optima increase by factors of 4--8 typically, moving from A to B or to C.  One might expect that the cases with larger numbers of local optima, and larger changes in the numbers of local optima when the higher modes are added, are more sensitive to small changes in body motions. One question is whether the optimal efficiencies in these cases (e.g. the values at the top left of figure \ref{fig:Mode1SymmOptimaParams}A) have larger increases when higher modes are added.

\begin{figure} [h]
           \begin{center}
           \begin{tabular}{c}
               \includegraphics[width=7in]{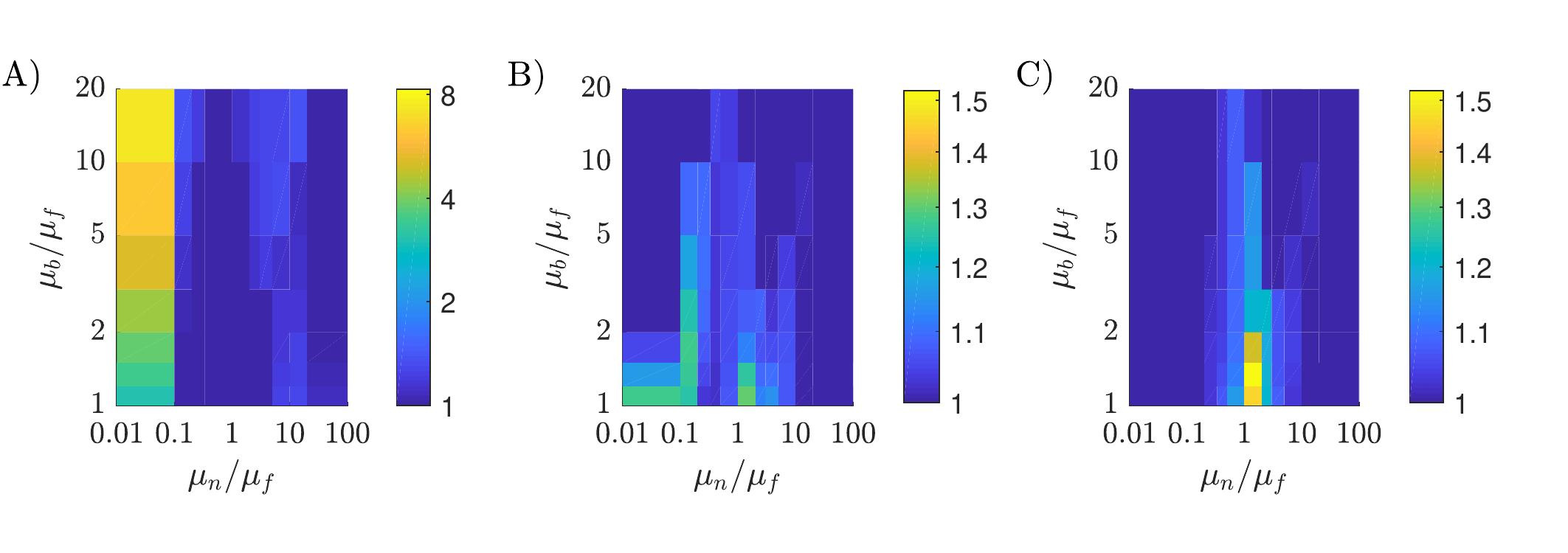} \\
           \vspace{-.25in} \hspace{-.25in}
           \end{tabular}
          \caption{\footnotesize The factor of improvement in maximum relative efficiency when the space of modes is enlarged from A) $\{A_1, B_1\}$ to $\{A_0, A_1, B_1\}$; B) $\{A_0, A_1, B_1\}$ to $\{A_0, A_1, B_1, A_2, B_2\}$; C) $\{A_0, A_1, B_1\}$ to $\{A_0, A_1, B_1, A_3, B_3\}$. The modes corresponding to these coefficients are listed in equations (\ref{BilatEllipseHigher}). The factor is plotted at various friction coefficient values shown on the axes. 
 \label{fig:ImprovementsFig}}
           \end{center}
         \vspace{-.10in}
        \end{figure}

Figure \ref{fig:ImprovementsFig} shows the changes in peak efficiency when the parameter space is enlarged from smaller to larger sets of modes in (\ref{BilatEllipseHigher}). First, we consider the improvement when motions that are asymmetric with respect to the flat state are considered, for elliptical trajectories. Panel A shows the factor of improvement in the peak efficiency when
modes with $\{A_0, A_1, B_1\}$ are considered compared to those with just $\{A_1, B_1\}$. At the smallest $\mu_n/\mu_f$, the $A_0$ term allows for a large increase the peak relative efficiency. At most other friction coefficient ratios, there is no improvement, except in a strip of values contained within $1 \leq \mu_n/\mu_f \leq 10$, where the improvement is typically 20--30\%. Panel B shows the improvement obtained by expanding from $\{A_0, A_1, B_1\}$ to $\{A_0, A_1, B_1, A_2, B_2\}$.  It is somewhat surprising that in most cases here, there is little improvement from considering these two additional modes. There is little to no improvement except near isotropic friction and near $0.01 \leq \mu_n/\mu_f \leq 0.1$ where the improvement is at most 31\%. Panel C shows the improvement from expanding from $\{A_0, A_1, B_1\}$ to $\{A_0, A_1, B_1, A_3, B_3\}$. Here too, the improvement is modest, with improvements up to 51\% near isotropic friction, but less than 7.5\% outside of $1/3 \leq \mu_n/\mu_f \leq 3$. Taken together, these results suggest that in most cases ellipses, in particular ellipses centered at the origin, may be good approximations to the optimal trajectories with large numbers of modes. Our stochastic optimization results shown later will support this statement, except in some cases with $\mu_n/\mu_f \ll 1$.

\begin{figure} [h]
           \begin{center}
           \begin{tabular}{c}
               \includegraphics[width=7in]{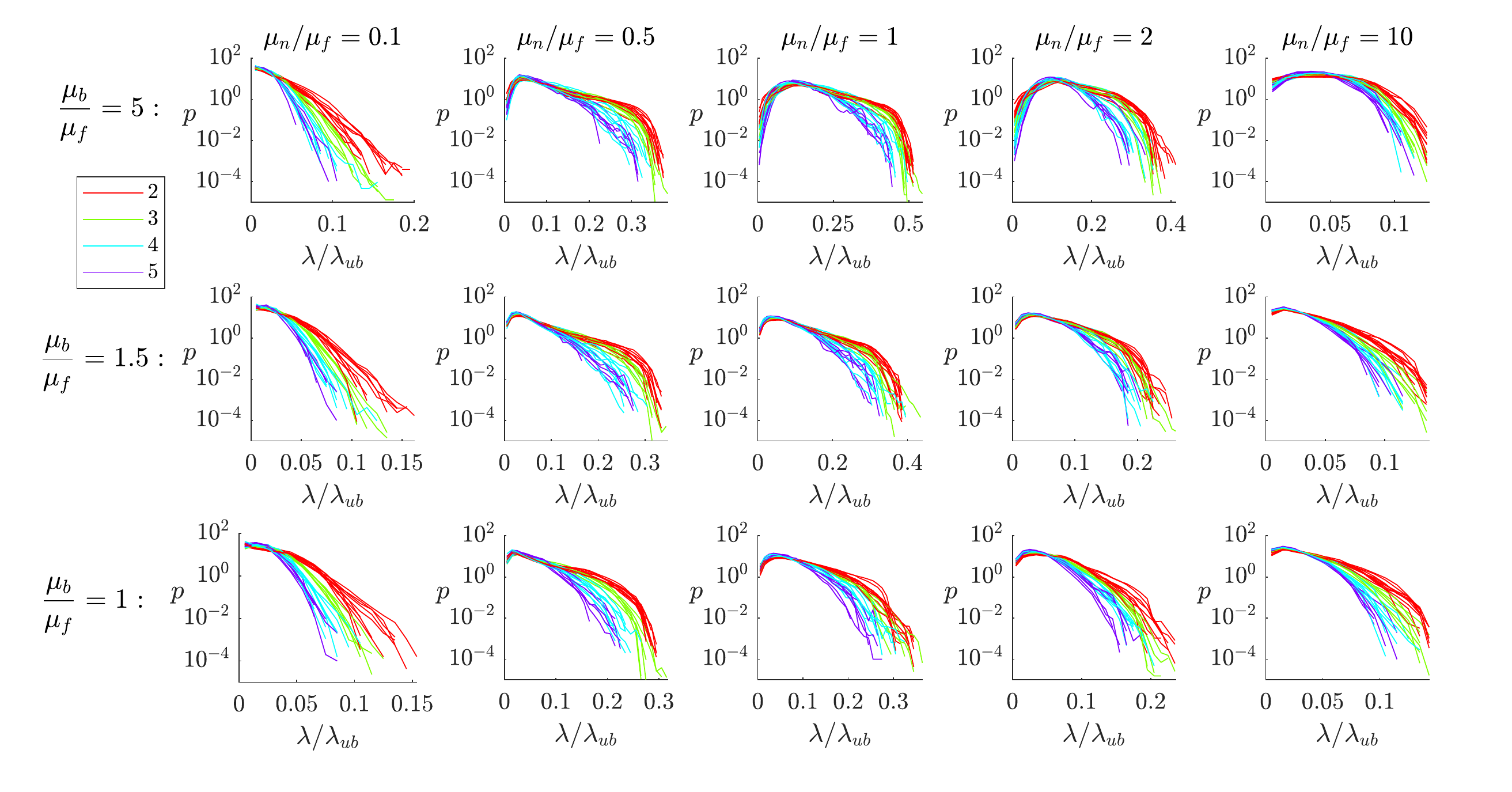} \\
           \vspace{-.25in} \hspace{-.25in}
           \end{tabular}
          \caption{\footnotesize Probability densities of relative efficiency, estimated from histogram data for various friction coefficient ratios (labeled at top and left). Each color corresponds to bilaterally symmetric trajectories with a given maximum frequency $k$, labeled at left, resulting in $2k+1$ modes. Each curve corresponds to a different random ensemble of about $10^6$ trajectories.  
 \label{fig:ModeRand0KOrigSymmOptima}}
           \end{center}
         \vspace{-.10in}
        \end{figure}

As the number of modes increases above five, it rapidly becomes prohibitively expensive to compute results across a grid that resolves all of the coefficient parameter space. We explore higher-dimensional spaces by instead selecting a random ensemble of $\approx 10^6$ points in coefficient space. For example, with frequencies up to $k$ = 3, there are seven coefficients in (\ref{BilatEllipseHigher}). A large ensemble of seven-component vectors is chosen, with each of the seven components (the coefficients) drawn from a uniform distribution on $[-1.2\pi, 1.2\pi]$. Most points yield trajectories that include self-intersecting bodies at certain times, and these are eliminated. The relative efficiency is computed for the non-self-intersecting cases, $\approx 10^6$ in number. This is done for $k$ = 2, 3, 4, and 5 frequencies, with coefficients in a $2k+1$-dimensional space, and ten different random ensembles in each case. For each ensemble, we bin the data in small increments of relative efficiency, and construct an estimate of the probability density of relative efficiency, plotted for each $k$ in figure \ref{fig:ModeRand0KOrigSymmOptima}, on the five-by-three grid of friction coefficient ratios used in figure \ref{fig:RotationDataFigure}. The densities typically have a peak at an efficiency that is some distance from zero (except in the leftmost column), the typical efficiency magnitude for a random kinematics. 
After the peak, the densities fall off exponentially (a linear behavior on this log-linear scale) or faster. They are many orders of magnitude smaller near the maximum efficiencies, which are therefore rare events. There is some scatter among the ten different random ensembles (the set of ten curves with the same color in each panel), particularly at the largest efficiencies. Nonetheless, the curves of a given color tend to cluster together, and near the peaks the densities are not very sensitive to the particular ensemble used. In most cases, $k = 2$ gives the best performance---the highest density of states at large efficiencies---and the performance decreases with larger $k$. The spaces with lower $k$ are nested in those at higher $k$, so the maximum efficiency over all kinematics must occur in the space with largest $k$. However, figure \ref{fig:ModeRand0KOrigSymmOptima} shows that it is unlikely to arise in the samples chosen. The method of sampling (uniform sampling in each coefficient, with self-intersecting motions discarded) could affect the increased prevalence of lower-efficiency states at larger $k$. For example, many kinematics with large high-frequency components may be ineffective for locomotion, and these are likely to occur with the uniform sampling of each coefficient used here.

\section{Stochastic optimization}

We have presented the relative efficiency for individual optima, their kinematics in the elliptical case, and some of the features of trajectory spaces---numbers of optima, distributions of rotations and efficiencies, and incremental improvements from enlarging the space of modes---with dimensions up to 11 (i.e. $k$ up to 5). We now study the features of optimal trajectories as the space of modes is increased further, by using a stochastic optimization method with ensembles of trajectories. Compared to the quasi-Newton approaches used in \cite{AlbenSnake2013,TaHo2007a}, the stochastic method is gradient-free, and therefore simpler to implement---particularly given the constraint that trajectories remain in the non-self-intersecting region. The main drawbacks are that more iterations are needed to obtain convergence, and the stochastic algorithm requires parameters that are tuned heuristically, unlike the more standardized Newton-type search algorithms \cite{pham2012intelligent}.

Here we create a large number of populations (e.g. 250), each population with 50 trajectories, and evolve the populations over many generations. At each generation, we evaluate the relative efficiency of each trajectory, select the top 50\%
of trajectories, and replace the entire population with random perturbations of the top 50\%. We add perturbations to the coefficients, drawn from uniform or Gaussian distributions. The magnitude of the perturbation is a tuned parameter, typically 0.001--0.01 multiplied by the reciprocal of the frequency corresponding to the coefficient. If the perturbation magnitude is at the smaller end of the range, the population converges slowly but directly to the nearest local optimum. If the perturbation magnitude is at the larger end, the population converges more quickly and possibly to a wider range of optima, but fluctuates more around a given optimum. Therefore, we start with a larger perturbation magnitude and progressively decrease it, as in simulated annealing \cite{pham2012intelligent}. We run each population for 1000 generations, by which point convergence is obtained. %With 250 populations there is the possibility of different populations converging to different optima. 

\begin{figure} [h]
           \begin{center}
           \begin{tabular}{c}
               \includegraphics[width=7in]{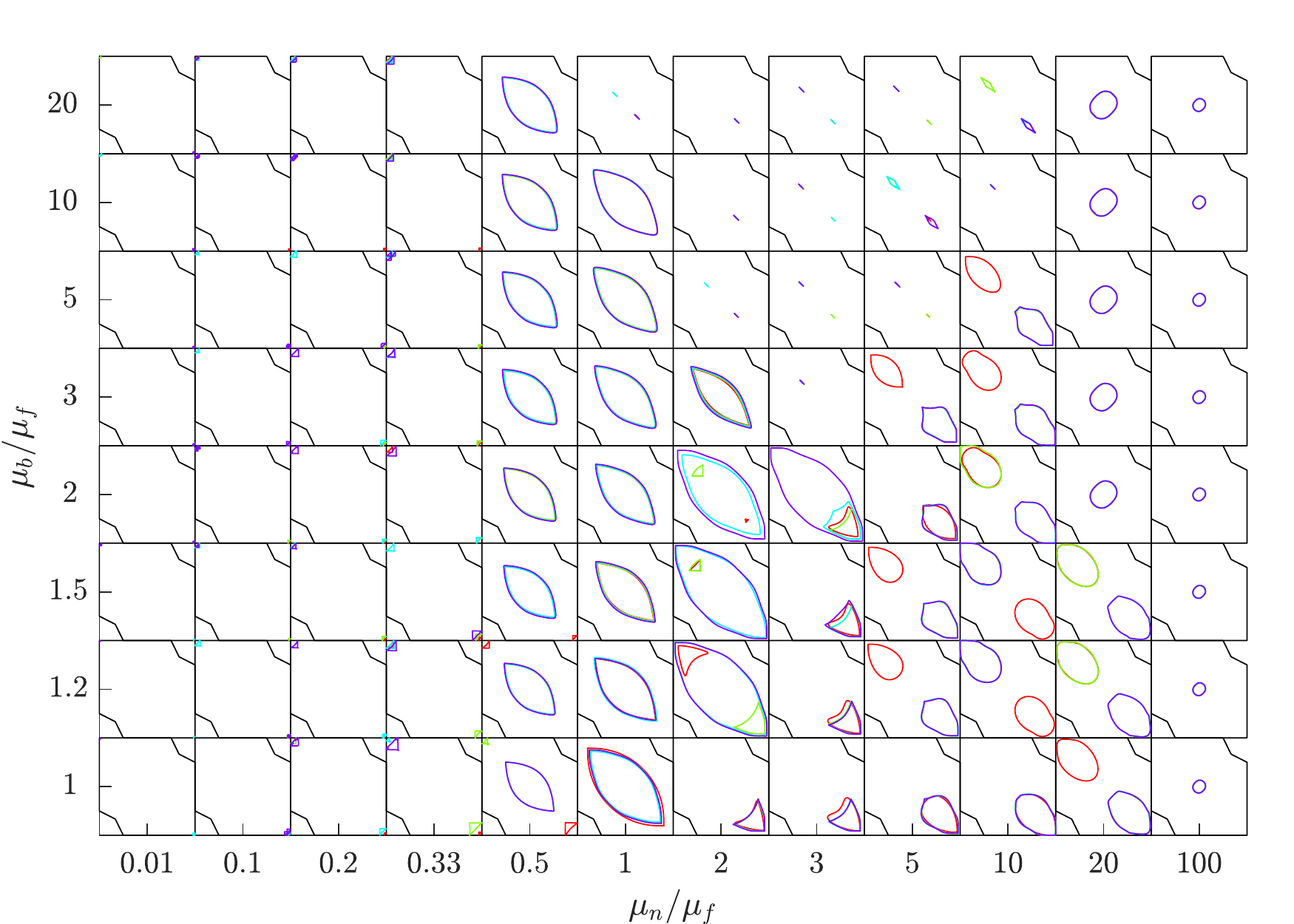} \\
           \vspace{-.25in} \hspace{-.25in}
           \end{tabular}
          \caption{\footnotesize Efficiency-maximizing trajectories with bilateral symmetry, with different maximum frequencies $k$---3 (red), 5 (green), 7 (light blue), and 9 (purple)---corresponding to 2$k$+1 modes in each case. The trajectories are plotted in the region of nonintersecting trajectories, plotted at various friction coefficient ratios labeled at bottom and left. The trajectories are computed with the stochastic optimization algorithm described in the text. 
 \label{fig:OptimalBilateralTrajectories}}
           \end{center}
         \vspace{-.10in}
        \end{figure}

In figure \ref{fig:OptimalBilateralTrajectories} we plot the optimal trajectory thus obtained, among all the populations, in friction coefficient space. The trajectories are plotted within the region of non-self-intersecting shapes, outlined in black at each pair of friction coefficient ratios. We consider trajectories with bilateral symmetry here. Different colors correspond to different maximum frequencies $k$ in (\ref{BilatEllipseHigher})---3 (red), 5 (green), 7 (light blue), and 9 (purple)---with 2$k$+1 modes in each case. As for the elliptical trajectory optima in figure \ref{fig:Mode1SymmOptimaMotions}, certain types of trajectories are strongly correlated with certain regions of friction coefficient space. There is generally very good agreement between the optima with different $k$. On the left, $\mu_n/\mu_f \ll 1$, the optimal trajectories are very small, in most cases almost 45-45-90 right triangles with two sides aligned with the $\Delta\theta_1$ and $\Delta\theta_2$ axes, and close to the upper left or lower right corners of each subregion. These are two versions of the same motion (symmetric about the line $\Delta\theta_1 =\Delta\theta_2$, i.e. with $\Delta\theta_1$ and $\Delta\theta_2$ interchanged), with the body executing very small motions about a mean shape than is nearly completely folded together as in motion 8 of figure \ref{fig:Mode1SymmOptimaMotions}C. The triangles are largest and easiest to see at $\mu_n/\mu_f = 0.33$ and $\mu_b/\mu_f = 1$, and gradually become smaller moving leftward and upward in friction coefficient space.
There is a transition to much larger lenticular or oval-shaped trajectories, centered at the origin at $\mu_n/\mu_f = 0.5$. These become larger, eventually filling the non-self-intersecting region at $\mu_n/\mu_f = 2$ for some $\mu_b/\mu_f$. Here and at $\mu_n/\mu_f = 3$, triangular trajectories in the corners reappear, this time more curved and larger than previously. For larger $\mu_b/\mu_f$ and $1 \leq \mu_n/\mu_f \leq 10$, small slit trajectories appear,  very similar to motion 9 in figure \ref{fig:Mode1SymmOptimaMotions}C, and occurring at similar friction coefficients. At smaller $\mu_b/\mu_f$, as $\mu_n/\mu_f$ ranges from 5 to 20, the corner trajectories become larger and rounder, and at the largest $\mu_n/\mu_f = 100$, all the trajectories become small circles at the origin, like kinematics 3 in figure \ref{fig:Mode1SymmOptimaMotions}C, but symmetric about the flat state, and moving mainly tangentially at this pair of friction coefficient ratios. Most of these trajectories are simple closed curves that can be approximated reasonably well by ellipses. 

%Note with Bilateral symm, lower mode optima is better at smaller mut, higher mode is slightly better at higher mut. Efficiency is sensitive to detailed shape at lower mut.

%With Antipodal symm, lower mode is better only at extremes, mut/muf very low, very high

\begin{figure} [h]
           \begin{center}
           \begin{tabular}{c}
               \includegraphics[width=7in]{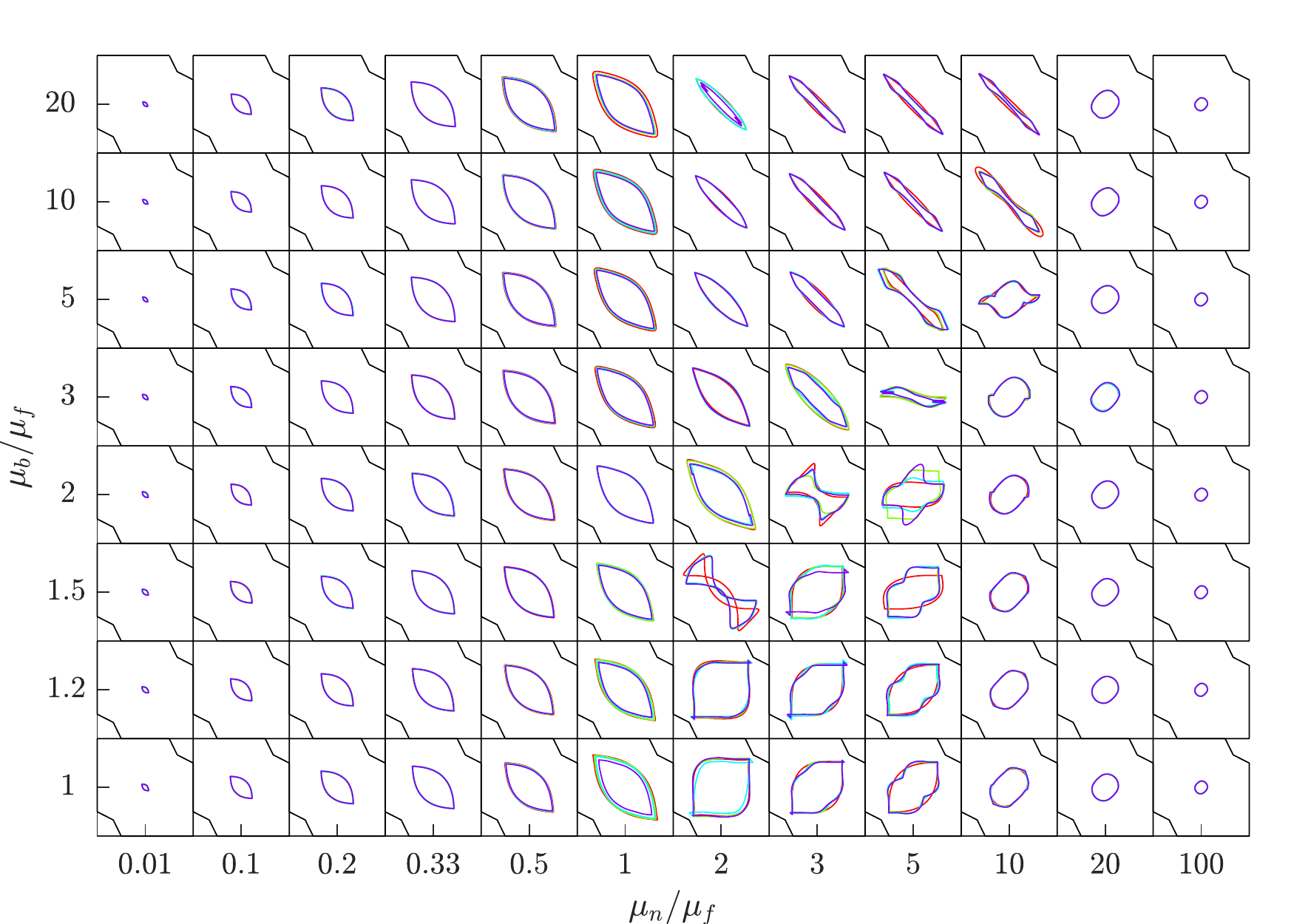} \\
           \vspace{-.25in} \hspace{-.25in}
           \end{tabular}
          \caption{\footnotesize  Efficiency-maximizing trajectories with antipodal symmetry, with different maximum frequencies $k$---3 (red), 5 (green), 7 (light blue), and 9 (purple)---corresponding to 2$k$+2 modes in each case. The trajectories are plotted in the region of nonintersecting trajectories, plotted at various friction coefficient ratios labeled at bottom and left. The trajectories are computed with the stochastic optimization algorithm described in the text.
 \label{fig:OptimalAntipodalTrajectories}}
           \end{center}
         \vspace{-.10in}
        \end{figure}

Figure \ref{fig:OptimalAntipodalTrajectories} shows the results with the same optimization procedure but for the other main class of zero-net-rotation trajectories---those with antipodal symmetry. The values of $k$ are the same, but result in 2$k$+2 modes now using (\ref{AntipEllipseHigher}). Except in a few cases (e.g. (5, 3), (10, 5)), these trajectories also have the bilateral symmetry of the previous trajectories. Where the trajectories in figure \ref{fig:OptimalBilateralTrajectories} are centered at the origin, the two types of optima agree well. Where they disagree, if the antipodally symmetric optima also have bilateral symmetry (as in nearly every case), they must be inferior, or else they would also occur in figure \ref{fig:OptimalBilateralTrajectories}. In general, the antipodally symmetric optima vary more smoothly in parameter space. 

\begin{figure} [h]
           \begin{center}
           \begin{tabular}{c}
               \includegraphics[width=7in]{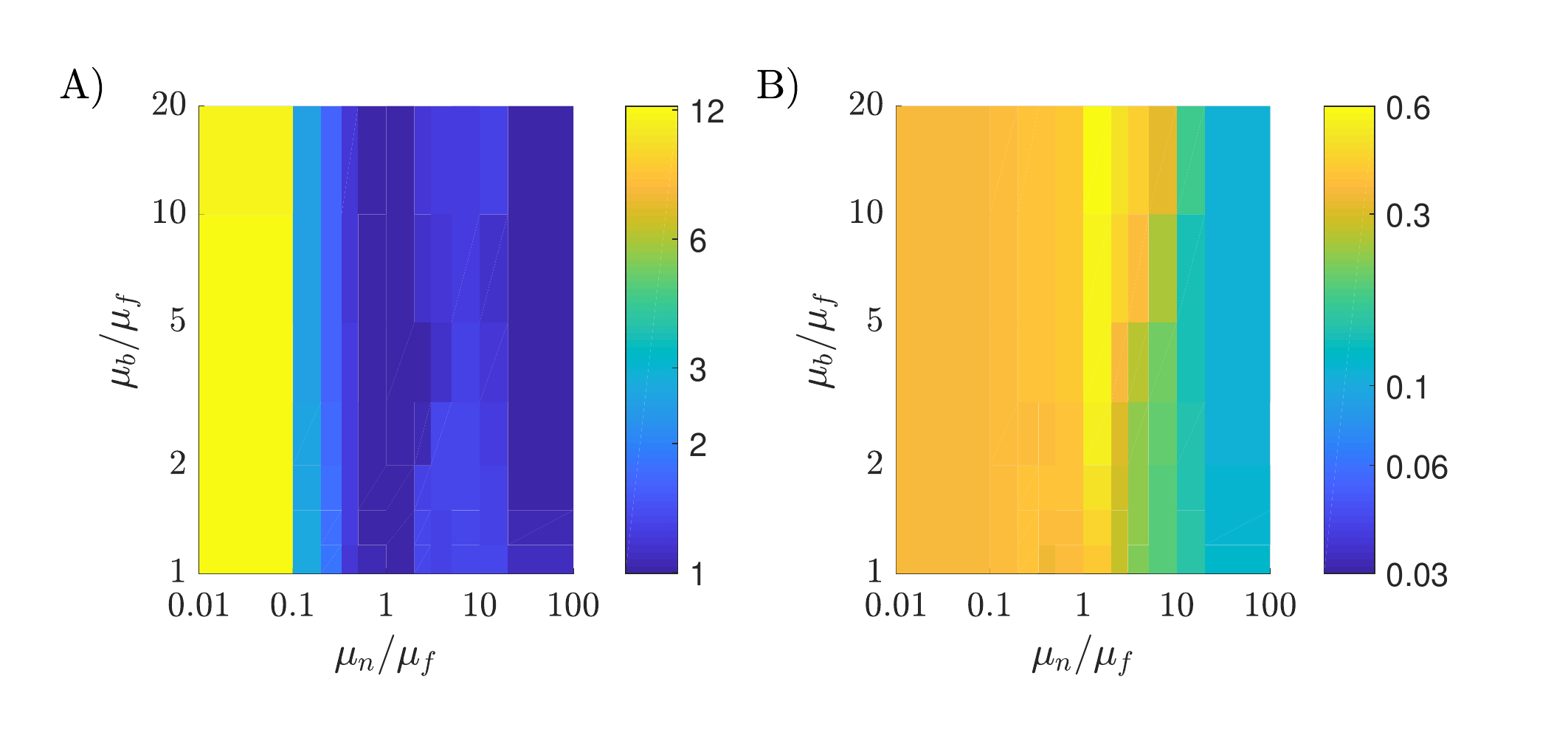} \\
           \vspace{-.25in} \hspace{-.25in}
           \end{tabular}
          \caption{\footnotesize A) The factors by which the efficiencies of the bilaterally symmetric optima exceed those of the antipodally symmetric optima. B) The relative efficiencies of the bilaterally symmetric optima. 
 \label{fig:BilateralAntipodeRatio}}
           \end{center}
         \vspace{-.10in}
        \end{figure}

For all friction coefficient ratios, we find that the optima with bilateral symmetry are as good as those with antipodal symmetry, and often much better. The factors by which the efficiencies of the bilaterally symmetric optima exceed those of the antipodally symmetric optima are plotted in figure \ref{fig:BilateralAntipodeRatio}A. The factor is about 12 at $\mu_n/\mu_f = 0.01$, 2.4--2.7 at $\mu_n/\mu_f = 0.1$, and decreases to about 1 at $\mu_n/\mu_f$ = 0.5 and 1. It then rises again to 1.2--1.3 for $2 \leq \mu_n/\mu_f \leq 10$, and then drops back to 1 for larger $\mu_n/\mu_f$. The values of the relative efficiency for the bilaterally symmetric optima are shown in panel B. They are fairly uniform, 0.34--0.42, on the left side of the panel, $0.01 \leq \mu_n/\mu_f \leq 0.5$. On the right side of the panel, they are similar to the values for the elliptical optima in figure \ref{fig:Mode1SymmOptimaParams}A, except near isotropic friction. There the bilaterally symmetric optima are about 60\% more efficient, but the advantage decreases rapidly moving to larger $\mu_b/\mu_f$ and $\mu_n/\mu_f$. This is consistent with the fact that the trajectories in figure \ref{fig:OptimalBilateralTrajectories} become either more rounded (at large $\mu_n/\mu_f$) or flat (at large $\mu_b/\mu_f$), in both cases closer to ellipses. For 
$\mu_n/\mu_f = 0.01$, the bilaterally symmetric optima are about six times as efficient as the elliptical optima. Here the efficiency is sensitive to the detailed shape of the trajectory (i.e. triangular versus flat), and the higher frequencies are needed to approximate the optimal trajectory for a three-link body.

\subsection{Linear resistance}

Many previous works have considered the optimal motions of three-link swimmers at zero Reynolds number \cite{Pu1977a,BeKoSt2003a,TaHo2007a,AvRa2008a,RaAv2008a,HaBuHoCh2011a,huber2011micro,passov2012dynamics,
alouges2013self,giraldi2015optimal,wiezel2016optimization,hatton2017kinematic,bettiol2017purcell}. To compare with this important case, we now consider how the optimal trajectories change when the resistive force is linear in velocity, instead of speed-independent as in the preceding results. This corresponds to resistive force theory, which is the lowest-order approximation to the nonlocal viscous forces on a slender body \cite{cox1970motion}. Although nonlocal slender body theories have also been developed and used extensively \cite{TaHo2007a,lauga2009hydrodynamics}, resistive force theory gives a sufficient representation of the physics for many swimming problems \cite{zhang2014effectiveness,peng2017propulsion,keaveny2017predicting,schnitzer2018resistive,lopez2020hydrodynamic}.  
The anisotropy ratio for a long cylinder, $\mu_n/\mu_f = 2$, has been used most often for a body swimming in a Newtonian fluid \cite{cox1970motion,BeKoSt2003a,lauga2009hydrodynamics}. \cite{cohen2010swimming} mentions a value of 1.5 as more appropriate for undulating bodies; \cite{riley2017empirical} mentions values between 1 and 2 in an empirical theory for shear-thinning fluids; and \cite{leshansky2009enhanced} derives ratios both less than and greater than two for complex fluids. Ratios greater than 2 (of the order of 10) have also been used to model the crawling of microorganisms on wet surfaces \cite{sauvage2011elasto,shen2012undulatory,rabets2014direct,keaveny2017predicting}. We are unaware of studies that derive ratios smaller than 1 for biological or physical swimmers, though \cite{taylor1952analysis,cohen2010swimming} mention the possibility for the marine worm Nereis, which have enhanced resistance along the body axis due to bristles, and use direct wave locomotion. We are also unaware of swimmers that have been modeled with $\mu_b/\mu_f$ different from 1, but some difference would occur with bodies that are not fore-aft symmetric. For comparison with the sliding locomotion results in this paper, we compute optimally efficient trajectories with the linear resistance law in the same space of ratios of resistance (previously friction) coefficients.

For the case of linear resistance, we replace $\widehat{\partial_t{\mathbf{X}}}_\delta$ by $\partial_t{\mathbf{X}}$ in (\ref{frictiondelta}). Bilaterally and antipodally symmetric trajectories still yield no net rotation; the cancellations in rotation are not affected by how the resistive force depends on the velocity magnitude. The definition of efficiency is changed from (\ref{lambda}) to $\lambda = \| \overline{\partial_t \mathbf{X}} \|^2/\langle P \rangle$ as in previous works \cite{BeKoSt2003a,TaHo2007a}, modulo constant factors. The same upper bound, $\lambda_{ub} = 1/\mu_{min}$, holds with resistance that is linear in velocity, as follows. We now have
\begin{align}
\langle P \rangle = \int_0^1 \int_0^1 \mu_n (\partial_t \mathbf{X} \cdot \hat{\mathbf{n}})^2 +
\left( \mu_f H(\partial_t \mathbf{X}\cdot \hat{\mathbf{s}})
+ \mu_b (1-H(\partial_t \mathbf{X}\cdot \hat{\mathbf{s}}))\right) (\partial_t \mathbf{X} \cdot \hat{\mathbf{s}})^2 ds \, dt \geq \mu_{min} \int_0^1 \int_0^1 
\| \partial_t \mathbf{X} \|^2 ds \, dt.
\end{align}
We decompose $\partial_t \mathbf{X}$ into its time-and-space average $\overline{\partial_t \mathbf{X}}$ plus the remainder $\widetilde{\partial_t \mathbf{X}}$, which has zero time-and-space average:
\begin{align}
\partial_t \mathbf{X} = \overline{\partial_t \mathbf{X}} + \widetilde{\partial_t \mathbf{X}} \qquad; \qquad
\overline{\partial_t \mathbf{X}} \equiv \int_0^1 \int_0^1 \partial_t \mathbf{X} ds \, dt.
\end{align}
Therefore
\begin{align}
\langle P \rangle \geq \mu_{min} \int_0^1 \int_0^1 
\| \partial_t \mathbf{X} \|^2 ds \, dt &= \mu_{min} \int_0^1 \int_0^1 
\| \overline{\partial_t \mathbf{X}} \|^2 + \| \widetilde{\partial_t \mathbf{X}} \|^2 ds \, dt + 2\mu_{min} \overline{\partial_t \mathbf{X}} \cdot \int_0^1 \int_0^1 \widetilde{\partial_t \mathbf{X}} \,   ds \,  dt\\
&=  \mu_{min} \int_0^1 \int_0^1 
\| \overline{\partial_t \mathbf{X}} \|^2 + \| \widetilde{\partial_t \mathbf{X}} \|^2 ds \, dt  \geq \mu_{min} \| \overline{\partial_t \mathbf{X}} \|^2.
\end{align}
Therefore, for a given average speed $\| \overline{\partial_t \mathbf{X}} \|$, $\langle P \rangle$ is at least $\mu_{min} \| \overline{\partial_t \mathbf{X}} \|^2$, which occurs
when all points of the body move uniformly
in the direction of minimum resistance, at constant speed $\| \overline{\partial_t \mathbf{X}} \|$. This provides the upper bound on efficiency:
\begin{align}
\lambda = \frac{\| \overline{\partial_t \mathbf{X}} \|^2}{\langle P \rangle} \leq \lambda_{ub} = \frac{1}{\mu_{min}}. 
\end{align}

\begin{figure} [h]
           \begin{center}
           \begin{tabular}{c}
               \includegraphics[width=7in]{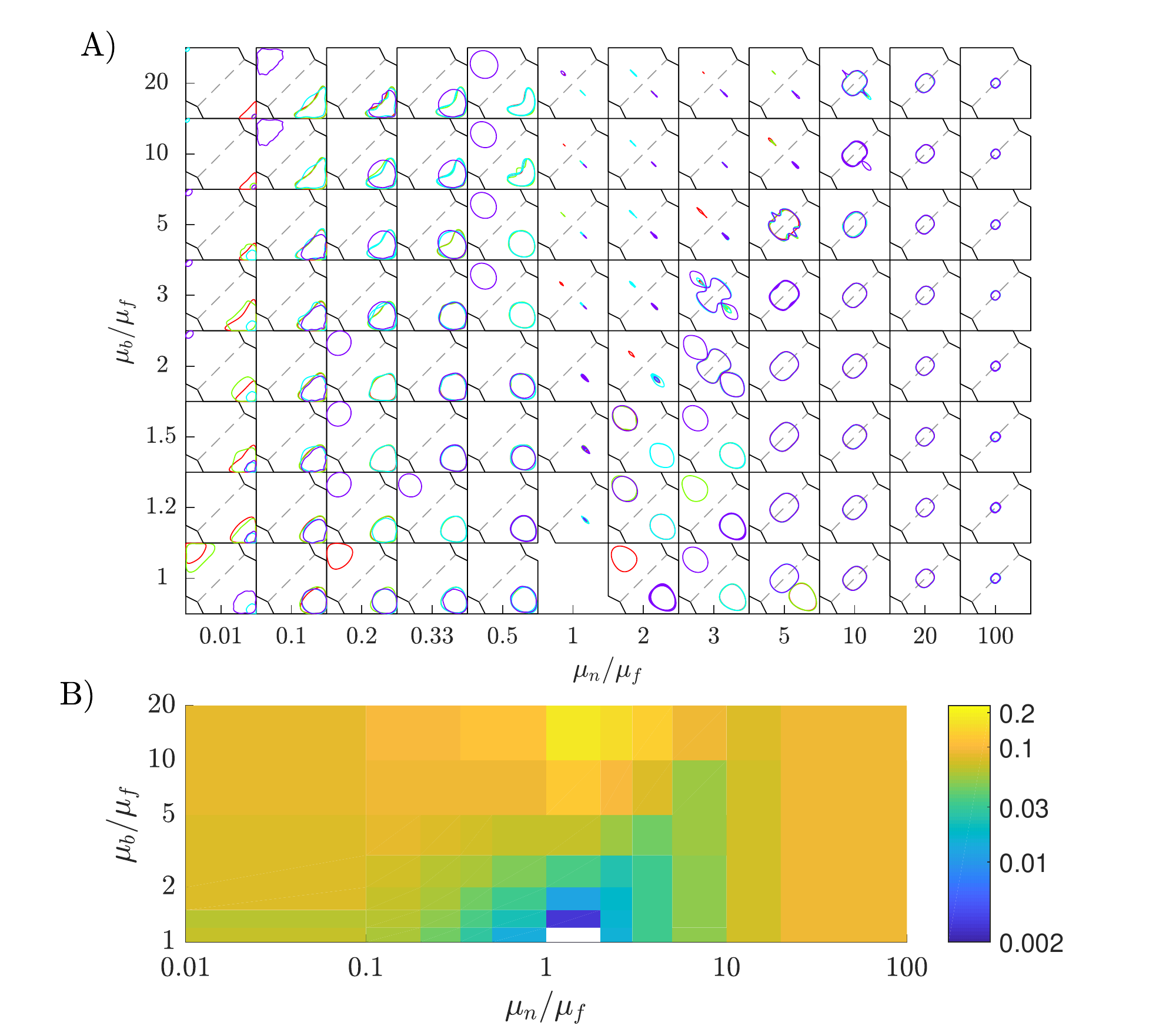} \\
           \vspace{-.25in} \hspace{-.25in}
           \end{tabular}
          \caption{\footnotesize  A) Trajectories (with bilateral or antipodal symmetry) that maximize relative efficiency, with different maximum frequencies $k$---3 (red), 5 (green), 7 (light blue), and 9 (purple)---when the resistance law is linear in velocity. B) The relative efficiencies corresponding to the motions in panel A.
 \label{fig:OptimalLinearMotions}}
           \end{center}
         \vspace{-.10in}
        \end{figure}

%Many previous studies have considered optimally efficient motions of three-link swimmers in viscous fluids \cite{Pu1977a,BeKoSt2003a,TaHo2007a,AvRa2008a,RaAv2008a,HaBuHoCh2011a,huber2011micro,passov2012dynamics,
%alouges2013self,giraldi2015optimal,wiezel2016optimization,hatton2017kinematic,bettiol2017purcell}.
%We now include a comparison of the previous optima with those that arise when the resistance law is linear in velocity.  
Figure \ref{fig:OptimalLinearMotions}A shows the trajectories (computed with the stochastic algorithm) that maximize relative efficiency, among the class of trajectories with either bilateral or antipodal symmetry, when the resistance law is linear in velocity. At large $\mu_n/\mu_f$, the trajectories are similar to those with Coulomb friction in figure \ref{fig:OptimalAntipodalTrajectories}. Near $\mu_n/\mu_f = 2$,
the trajectories are off-center, like those in figure \ref{fig:OptimalBilateralTrajectories}, and like that proposed by
\cite{RaAv2008a} for high efficiency, but those in figure \ref{fig:OptimalLinearMotions} are rounder. At 
$\mu_n/\mu_f = \mu_b/\mu_f = 1$, all trajectories yield zero locomotion with linear resistance \cite{alben2019efficient}, so none is shown.
For  $\mu_b/\mu_f > 1$ and $\mu_n/\mu_f$ = 1 and somewhat larger, small-amplitude reciprocal motions are optimal,
similar to those in the sliding case, figure \ref{fig:OptimalBilateralTrajectories}. The symmetrical lenticular or oval shapes 
in the central parts of figures \ref{fig:OptimalBilateralTrajectories} and \ref{fig:OptimalAntipodalTrajectories} do not appear in figure \ref{fig:OptimalLinearMotions}. Here, rounded off-center trajectories appear at both $\mu_n/\mu_f > 1$
and $< 1$. Decreasing $\mu_n/\mu_f$ to 0.1 and with $\mu_b/\mu_f > 1.5$, the trajectories become somewhat triangular, and very small in size as $\mu_n/\mu_f$ is decreased further to 0.01, roughly like those in figure \ref{fig:OptimalBilateralTrajectories}, but not as small. In general, many of the optimal trajectories with linear resistance resemble those with the Coulomb friction resistance law. The differences are most pronounced in the vicinity of isotropic friction, where linear resistance yields no locomotion. Figure \ref{fig:OptimalLinearMotions}B shows the distribution of optimal relative efficiencies corresponding to panel A. The distribution is similar to that of figure \ref{fig:BilateralAntipodeRatio}B. The maximum occurs at the top center in both cases. Value decrease moving leftward and rightward, more to the right in figure \ref{fig:BilateralAntipodeRatio}B but more symmetrically in figure \ref{fig:OptimalLinearMotions}B. The relative efficiency values are generally much lower for linear resistance---about 0--30\% of the values for Coulomb friction in the left half of figure \ref{fig:BilateralAntipodeRatio}B, $\mu_n/\mu_f < 1$. In the right half, they are also generally much lower, but reach 50\% of the Coulomb friction values when $\mu_n/\mu_f$ increases to 10, and exceed the Coulomb friction values by a few percent along the rightmost boundary, $\mu_n/\mu_f = 100$. 

\section{Summary and conclusions}

We have investigated efficient sliding motions of three-link bodies with a Coulomb friction resistance law and various frictional anisotropy ratios. We found that the reduced space of elliptical (single-frequency) trajectories gives a good representation of optimal motions when more frequencies are considered. Friction coefficient space can be partitioned into distinct regions (about seven are suggested here for elliptical trajectories) where different types of motions are optimal. Surprisingly, the top two elliptical optima usually belong to different clusters in trajectory coefficient space, despite having similar relative efficiencies, showing that very different motions can be close to optimal for a given choice of friction anisotropy ratios. Many of the elliptical optima bend symmetrically to either side of the flat state, but several optima are strongly asymmetrical, including small-amplitude reciprocal motions. Some of the optima resemble those seen previously in the smooth case---small-amplitude retrograde or direct wave locomotion with very large or very small normal friction, reciprocal (or ratcheting) motions with large backward friction. But most of the optima are distinct from those seen previously. 

The elliptical motions with zero net rotation belong to three groups: those with a certain bilateral symmetry, antipodal symmetry, and a small subset of the reciprocal motions. For trajectories with frequencies up to a given multiple of the basic frequency, the first two groups have about half the dimension of general trajectories, but achieve about the same maximum efficiency, with a noticeable reduction only for very small normal friction. 

Adding modes with two or three times the basic frequency to bilaterally symmetric elliptical trajectories increases the number of local optima by a factor of 4--8 in most of parameter space, but increasing to about 20 when normal friction is small. Adding these modes increases the maximum efficiency by at most 50\%, and usually much less. We then considered random ensembles with uniformly distributed coefficients of modes that range up to five times the basic frequency. The probability density of efficiency has a peak at a nonzero efficiency in most cases, and falls off exponentially or faster up to the maximum efficiency value.  Ensembles that include higher frequencies are skewed towards smaller efficiencies.

We developed a stochastic optimization method to find optimal trajectories in larger spaces of modes, with up to nine times the basic frequency. We find rapid convergence with increasing numbers of modes. Bilaterally symmetric optima outperform antipodally symmetric optima where they differ. With small normal friction, the higher-frequency optima have the same general size and location as the elliptical optima, but have a triangular shape that increases efficiency by a factor as large as six at the smallest normal friction studied. At intermediate normal friction, the higher-frequency optima are similar to the elliptical optima, though perhaps with angular shapes, and efficiencies are only moderately higher. In nearly all cases, the higher-frequency optima are simple closed curves, even though simple closed curves are a small subset of the full set of trajectories (including self-intersection). With a linear resistance law, the peak relative efficiencies are much reduced, particularly near isotropic resistance where the efficiency is always zero. The optimal trajectories are similar to the Coulomb friction case at large normal friction, more off-center and rounded at moderate normal friction, and larger and more rounded triangular trajectories at very small normal friction. As with Coulomb friction, nearly reciprocal motions with very small amplitude predominate at large backward friction and moderate normal friction.

\begin{acknowledgments}
This research was supported by the NSF Mathematical Biology program under
award number DMS-1811889.
\end{acknowledgments}

\bibliographystyle{unsrt}
\bibliography{snake}

\end{document}